%% file: paper.tex
\documentclass{article}

\usepackage{bbold,enumerate}

\usepackage[curve,matrix,arrow,frame]{xy}
\CompileMatrices

\usepackage{diagrams}

\usepackage{latexsym}

\newtheorem{proposition}{Proposition}
\newenvironment{proof}{\noindent{\em Proof:}}{\hfill$\Box$\par}
\newcommand\termdef[1]{{\em #1}}
 
\newtheorem{DEFINITION}{Definition}

\newcommand\boxedvar[1]{*+<1.5em>[F-]{#1}}

\newcommand\N{{\mathbb N}}
\newcommand\refl{-}

\newcommand\fb[2]{{\hbox{\rm\texttt{fb}}}_{#1}({#2})}
\newcommand\bind[2]{{#1}\cdot{#2}}
\newcommand\product[2]{{#1}\times{#2}}
\newcommand\compose[2]{{#1}\bullet{#2}}
\newcommand\compto{\Rightarrow}
\newcommand\reachable[1]{\overline{#1}}
\newcommand\simto{\leadsto}

\newcommand\ack{{\tt ack}}
\newcommand\lock{{\tt lock}}
\newcommand\al{{\tt l}}
\newcommand\unlock{{\tt unlock}}
\newcommand\au{{\tt u}}
\newcommand\kbegin{{\tt begin}}
\newcommand\kend{{\tt end}}
\newcommand\kgo{{\tt go}}

\newcounter{robbiefoot}
\newcounter{giuliofoot}

\title{On Automata with Boundary}
\author{
R. Gates\footnote{Part of this work completed at
the University of Sydney.}\setcounter{robbiefoot}{\value{footnote}},\\
{\em\small Department of Computing, Division of ICS,}\\
{\em\small Macquarie University, N.S.W. 2109, Australia}\\
P. Katis\mbox{$^{\fnsymbol{robbiefoot}}$}
\footnote{This work has been supported by the Australian Research
Council. Progetto cofinanziato MURST ``Tecniche formali per la
specifica, l'analisi, la verifica, la sintesi e la
transformazione di sistemi software''.}
\setcounter{giuliofoot}{\value{footnote}},\\
{\em\small Dipartimento Scienze Chimiche, Fisiche e Matematiche,}\\
{\em\small Universit\`a degli Studi dell'Insubria}\\
N. Sabadini\mbox{$^{\fnsymbol{giuliofoot}}$},\\
{\em\small Dipartimento Scienze Chimiche, Fisiche e Matematiche,}\\
{\em\small Universit\`a degli Studi dell'Insubria}\\
R.F.C. Walters\mbox{$^{\fnsymbol{robbiefoot}}$}\mbox{$^{\fnsymbol{giuliofoot}}$}\\
{\em\small Dipartimento Scienze Chimiche, Fisiche e Matematiche,}\\
{\em\small Universit\`a degli Studi dell'Insubria}\\
}

\begin{document}
\maketitle

\begin{abstract}
\input{abstract}
\end{abstract}

\section{Introduction}
\label{section:intro}
\input{intro}

\section{Automata with Boundary}
\label{section:awb}
\input{awb}

\section{Model Checking}
\label{section:model}
\input{model}

\section{Simulation}
\label{section:simulation}
\input{sim}

\section{Further Examples}
\label{section:examples}
\input{examples}

\section{Conclusions and Future Directions}
\label{section:conclusion}
\input{conclusion}

\bibliographystyle{plain}
\bibliography{general}

\end{document}

%% file: abstract.tex
We present a theory of automata with boundary for designing, modelling and
analysing distributed systems.  Notions of behaviour, design and simulation
appropriate to the theory are defined.  The problem of model checking for
deadlock detection is discussed, and an algorithm for state space reduction
in exhaustive search, based on the theory presented here, is described.
Three examples of the application of the theory are given, one in the
course of the development of the ideas and two as illustrative examples of
the use of the theory.

%% file: intro.tex
In this paper, we shall present an introduction to and overview of the theory
of automata with boundary -- a transition system based approach to the
problem of designing, modelling and analyzing distributed systems.
The treatment explicitly models the boundaries between subsystems, across
which they communicate when part of a larger system. We describe notions of
comparison and simulation of automata that allow abstractions to be
performed in a compositional manner.  We give a notion of behaviour for
these automata which works fluidly with the operations used to construct
systems and the simulations used to abstract systems.  Finally, we observe
that this approach explicitly captures the design of a system as an element
of the theory, and we describe some of the advantages of this.

In section~\ref{section:awb}, we present a general introduction to automata
with boundary, behaviour, the operations used to construct systems from
subsystems, and the notion of a design.  This section develops the example
of the dining philosophers in conjunction with the theory to show the reader
how this well known example is treated using the ideas of this paper.

In section~\ref{section:model}, we address the problem of model checking,
specifically focusing on deadlock detection. We describe an algorithm (the
minimal introspective subsystem algorithm) which utilises the design
information for the system to assist in reducing the search
space required for exhaustive model checking techniques. The section
concludes with a description of the behaviour of the algorithm when
applied to the dining philosophers example.

In section~\ref{section:simulation}, we define comparison and simulation
of automata, and describe the sense in which they are compositional, and
the connection with behaviour and deadlock detection. We also give examples
of simulations related to the dining philosophers, and indicate how simulations
can be used to assist model checking, or to theoretically analyze systems of
interest.

In section~\ref{section:examples}, we give two further examples of the
theory -- the design of a simple scheduler, and a message acknowledgement
protocol. This section further illustrates the use of the theory and
indicates the range of applicability.

\medskip

The underlying mathematical formalism of the approach is that of category
theory (see~\cite{ls:conceptual} or~\cite{walters:CCS}) and more specifically
Cartesian bicategories (see~\cite{CW:cart_bicat}).
We stress, however, that no background in category theory is required in
this paper -- all the required definitions and results are stated in
terms of transition systems,
although we occasionally indicate in passing some key connections.

\bigskip

This paper forms part of
an ongoing research project to develop a compositional
theory of distributed systems using categorical structures.  For an overview
of the work of the project and the breadth of interpretation of
`distributed system', see~\cite{KSW:alg_feedback}.
The material in the current paper stems from work on
the bicategory Span(Graph) : in \cite{KSW:span_graph}, this
bicategory was shown to have precise connections with the algebra
of transition systems of Arnold and Nivat (\cite{arnold:transitions}),
and with the process algebras of Hoare
(\cite{hoare:comm_seq_proc}); and, in \cite{KSW:petri_span},
this bicategory was shown to be expressive enough to model
place/transition nets (\cite{thiagarajan:petri}).

%% file: awb.tex
The goal of this section is to present {\em automata with boundary} as
a model for systems composed from a number of communicating parts.
The {\em boundaries} form an integral part of our theory -- all interactions
of a system with its environment occur across its boundaries. We
discuss {\em behaviours} of an automaton, and how these behaviours
appear on boundaries of the automaton. Requirements on automata
can be expressed by restricting the behaviours of an automaton as they
appear on given boundaries.  We describe the operations {\em bind},
{\em feedback}, and {\em product} of automata with boundary, enabling
larger systems to be built by combining smaller systems in a way that
interacts well with behaviours. Importantly, these operations can
be represented pictorially as {\em designs}, allowing us to give
high level views of systems.

As we introduce the basic definitions, we shall present an example
illustrating the concepts being defined.  We shall use the example
of the dining philosophers, much favoured amongst works on distributed
systems.

\subsection{Automata with Boundary}
\label{section:sys_bdry}

The use of finite state automata to model transition systems has
a long history (see~\cite{arnold:transitions}). In this paper,
an automaton will consist of a
reflexive graph plus other data.  A \termdef{reflexive graph} consists
of a directed graph (with parallel edges and loops allowed), together
with a specified \termdef{reflexive edge}~$v \to v$ for each
vertex~$v$ of the graph. We do not insist that our automata be
finite, but all the examples we present are finite. We do restrict
our attention to finite automata when considering model checking.

A crucial element of our theory is that of {\em boundary}.  All
boundaries are typed by the kind of synchronization actions which
can occur across the boundary.  By an \termdef{action set}
we mean a finite set~$X$ with a distinguished element, denoted~$\refl$.
We refer to elements of~$X$ as \termdef{actions},
and~$\refl$ as the {\em trivial}, or {\em reflexive} action.

An \termdef{automaton with boundary}~$(S,(X_i,\mu_i)_{i \in I})$ consists
of the following data:
\begin{enumerate}
\item
	A reflexive graph~$S$, called the \termdef{state space} of the
	automaton, whose vertices are termed \termdef{states} and
	whose edges are termed \termdef{motions}.
\item
	A finite set~$I$ indexing the boundaries of the automaton.
\item
	For each~$i \in I$, a \termdef{boundary}~$(X_i, \mu_i)$ consisting
	of an action set~$X_i$ and a labelling~$\mu_i(e)$ of each
	motion~$e$ by an element of~$X_i$, such that~$\mu_i(e)$ is trivial
	if~$e$ is reflexive.
\end{enumerate}
Note that the state space of the automaton is the reflexive graph~$S$ --
it includes not only the states but the motions of the automaton, which
provide the cohesion of the states to justify the terminology of a space.
We have in mind that the reflexive edges of the state space~$S$ are
\termdef{idling} motions - see the comments after the definition of
simulation (section~\ref{section:simulations})
for a further exploration of this view.
 
The labelling of the motions indicates the
actions on the boundaries which accompany given motions, and we require
that if the automaton is idling, then it is idling on each boundary.  Of course,
nontrivial motions (i.e., motions which are not reflexive), may still
idle on some or all boundaries. Those motions idling on all boundaries
are called \termdef{internal}
motions.  They reflect the ability of an automaton to change state without
this being reflected in its interaction with the environment.

We shall occasionally say a motion of~$S$
\termdef{performs} an action on a boundary to mean that it is labelled
by the specified action on the specified boundary. Note also that
while a boundary consists both of the action set~$X$ and the labelling of
motions~$\mu$, we shall often speak of the boundary~$X$ when no
confusion arises. We shall say the \termdef{type} of a boundary~$(X,\mu)$
to mean the action set~$X$.

It is also worth noting that to give an action set is precisely to
give a reflexive graph with one vertex. That is, the action set may
be considered to be an automaton with trivial state.  While not explored
in this paper, a more general theory of this kind can relax this restriction,
and allow one to calculate with boundaries which possess internal state.

\subsubsection{Two boundary Automata}
\label{section:two_boundary_automata}

When dealing with an automaton, we typically focus temporarily on a
subfamily of the boundaries over which some operation is being performed --
for example the gluing of boundaries (i.e., {\em binding} -- see
section~\ref{section:bind} below).  Given a subset~$J$
of the set~$I$ indexing the boundaries of~$S$, we may write~$S$ as an
arrow
\[
\prod_{j \in J} X_j \rTo^{S} \prod_{k \in I\setminus J} X_k
\]
and picture the motions of~$S$ as being labelled in the two products via
tupling of the labelling on individual boundaries. We shall typically
abbreviate to just~$S \colon X \to Y$ when we wish to emphasize the division,
rather than the particular boundaries.  In this case, we shall term~$X$ the
\termdef{left} boundary, and~$Y$ the \termdef{right} boundary, of~$S$. The
passage from the product of~$X_i$'s to the single object~$X$ may be seen as
the collection of a bundle of wires into a single cable for purposes of
hierarchical design.  This view gives the connection between automata
with boundary and the theory of bicategories (see \cite{KSW:bicat_processes}).

This passage between multiple boundary and two boundary automata allows
for a more natural and workable definition of the operations on automata,
without sacrificing either expressive power or precision.

\subsubsection{Pictorial representation of Automata}
\label{section:pictures}

When describing automata, we typically draw pictures
by drawing the state space of the automaton, with edges labelled
to indicate the actions performed by the motions.
For an automaton~$S \colon X \to Y$, we write
the label~$(x|y)$ to indicate the motion performs the action~$x$ on the
left boundary and the action~$y$ on the right boundary.
We shall omit drawing reflexive edges, as they add no information to the
picture -- however the existence of these edges is crucial,
as they allow subautomata in bound systems to act independently
(see section~\ref{section:bind}, below).

We can depict automata with other than two boundaries in a similar manner.
For an automaton with boundaries indexed by~$I$, the labels on motions
are~\hbox{$I$-tuples} with entries drawn from the types of the corresponding
boundary.
In this case, we make explicit the correspondence between tuple entries and
boundaries.

\subsubsection{Automata for the Dining Philosophers}
\label{section:dp_automata}

For the example of the dining philosophers, we shall present two
automata with boundary -- a philosopher and a fork.  Each philosopher will
have two boundaries (the left and the right fork from her perspective),
and each fork will likewise have two boundaries (the philosophers who
can manipulate the fork). We shall thus have an action set~$L$, and
and automata with boundary~$P \colon L \to L$ (a philosopher)
and~$Q \colon L \to L$ (a fork).

The action set~$L$ consists of the actions which a philosopher and
a fork jointly perform. A given nontrivial action on which a
philosopher and a fork synchronize consists of the fork being either
picked up or put down.  We model this by taking the action
set~$L=\{\refl,\lock,\unlock\}$.

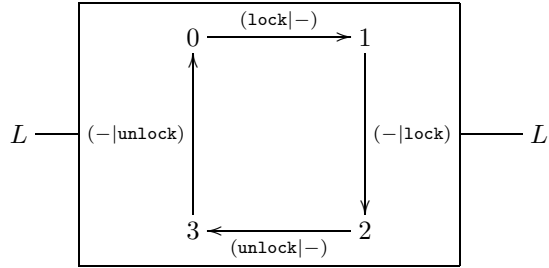
\begin{figure}
\[
\xymatrix
{
&& 0 \ar[rr]^{(\lock|-)}="top" && 1 \ar[dd]^{(-|\lock)}="right" \\
L && && && L \\
&& 3 \ar[uu]^{(-|\unlock)}="left" && 2 \ar[ll]^{(\unlock|-)}="bot"
\save
	"2,5"."left"."top"."bot"."right"*[F-]\frm{} \ar@{-}"2,1" \ar@{-}"2,7"
\restore
}
\]
\caption{\label{fig:phil} A Philosopher~$P$}
\end{figure}

The philosopher~$P$ is shown in figure~\ref{fig:phil}.
The philosopher has four states, corresponding to whether she is
attempting to acquire her left fork, acquire her right fork, relinquish
her left fork, or relinquish her right fork.  The motions
between these states are labelled by the boundary actions performed
by the motion.

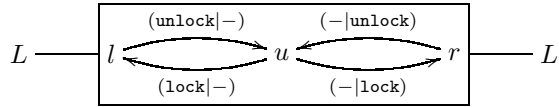
\begin{figure}
\[
\xymatrix
{
L & l \ar@/^/[rr]^{(\unlock|-)}="top" &&
	u \ar@/^/[ll]^{(\lock|-)}="bot" \ar@/_/[rr]_{(-|\lock)} &&
	r \ar@/_/[ll]_{(-|\unlock)} & L
\save
	"1,2"."top"."bot"."1,6"*[F-]\frm{} \ar@{-}"1,1" \ar@{-}"1,7"
\restore
}
\]
\caption{\label{fig:fork} A Fork~$Q$}
\end{figure}

The fork~$Q$ is shown in figure~\ref{fig:fork}. The fork has
three states, corresponding to whether it is unacquired (state~$u$),
acquired by its left boundary~(state~$l$), or acquired by its right
boundary~(state~$r$).  Once again, the motions are labelled by
the boundary actions they perform.

\subsection{Behaviour of Automata}
\label{section:behaviour}

For an automaton~$S \colon X \to Y$ and a state~$v_0$ of~$S$,
a \termdef{behaviour~$\beta$ of~$S$ with initial state~$v_0$} is
a sequence of motions~$(e_0, e_1, \ldots)$ (finite or infinite)
of~$S$ such that~$s(e_0) = v_0$ and~$s(e_{k+1}) = t(e_k)$ (for all
appropriate~$k$).
That is, a behaviour of~$S$ is nothing more than a path in its state
space. We write such behaviours
\[
\beta = v_0 \rTo^{e_0} v_1 \rTo^{e_1} \ldots
\]

Given a boundary~$(X,\mu)$ of~$S$, any behaviour of~$S$ is reflected
on the boundary via~$\mu$. Precisely, by the \termdef{appearance}
of a behaviour~$\beta = (e_0, e_1, \ldots)$ on~$X$ we mean the
sequence of actions~$(\mu(e_0), \mu(e_1), \ldots)$. We say a
\termdef{behaviour of~$S$ on a boundary~$X$} to mean a sequence which is
the appearance of some behaviour of~$S$ on~$X$.  Finally, we shall
refer to \termdef{reduced} appearances and behaviours on boundaries
to mean sequences obtained by eliding all trivial actions from
an appearance or behaviour.
It is crucial to note that a reduced appearance or behaviour need not
be an actual appearance or behaviour, as nontrivial motions of~$S$
may appear to be trivial actions on a given boundary.

An automaton gives rise to a relation between behaviours
on its boundaries.  Given a behaviour of~$S$ on the boundary~$X$ and
a behaviour of~$S$ on the boundary~$Y$, we may say these are related
if they are the appearances of the same behaviour of~$S$ on the boundaries
in question. It is
typical to specify a system by requesting this be a specific relation,
or by requesting properties of this relation. For example, performing
a given set of actions on the keypad of an automatic teller machine (one
of its boundaries) is required to result in cash being dispensed (the
action performed by the machine on the boundary with the cash dispenser)
and in an amount being deducted from the user's account (the action performed
by the machine on its boundary with the bank's account record).

For an automaton~$S \colon X \to Y$ and a given state~$v$ of~$S$, by the
\termdef{subautomaton of~$S$ reachable from~$v$} we mean the
automaton~$S' \colon X \to Y$ with states those states~$v'$ of~$S$ such
that exists a behaviour of~$S$ of the form
\[
v \rTo \ldots \rTo v'
\]
That is to say, there is a path of motions of~$S$ from~$v$ to~$v'$.
The motions of~$S'$ are all motions of~$S$ between states
of~$S'$, and the boundaries and labelling of~$S'$ are inherited directly
from~$S$.

We shall often assign to an automaton~$S$ of interest an initial state --
that is, a specified state~$v_0$ of~$S$.  In this case we speak
merely of \termdef{behaviours of~$S$} to mean behaviours of~$S$ with
initial state~$v_0$, and the \termdef{reachable subautomaton} to mean
the subautomaton reachable from~$v_0$.

The interpretation of time implied by this treatment of behaviour
is that of `discretized continuous time' -- there is an underlying
continuous time which is being discretely approximated by some fixed
time interval.  An aspect of this continuity is contained in the use
of the motions - each motion represents an atomic transition with
the same duration. We distinguish this from a purely `discrete time',
in which the motions are atomic processes which may have different
durations.

Thus while synchronization may be viewed as a real world process which
takes variable time - we model an instance of such a synchronization by
a behaviour consisting of internal motions bracketed by atomic
synchronizing transitions of the same duration.

\subsubsection{Behaviour for the Dining Philosophers}
\label{section:dp_behaviour}

Returning to the dining philosopher example of
section~\ref{section:dp_automata}, we choose the state~$0$
of the philosopher (figure~\ref{fig:phil}) to be the initial state.
A behaviour of the philosopher then consists of a repeating
sequence of the cycle ``\lock\ left boundary'', ``\lock\ right boundary'',
``\unlock\ left boundary, ``\unlock\ right boundary'', possibly
interspersed with reflexive edges.  The reduced appearance on a given
boundary is simply an alternating sequence of ``\lock'',``\unlock''
actions.

With the state~$u$ as initial, a behaviour of the fork
(figure~\ref{fig:fork}) consists of a sequence
of ``\lock\ boundary'', ``\unlock\ boundary'' pairs, with the boundary
possibly differing from pair to pair, and again possibly interspersed
with reflexive edges. Again, the reduced appearance on a given boundary
is an alternating sequence of ``\lock'',``\unlock'' actions.

\subsection{Operations: Binding, Feedback and Product}
\label{section:operations}

We describe three operations which may be used to construct new
automata with boundary from old.  In each case, we have a diagrammatic
view of the operation, which should be considered to be a {\em design} --
an expression in variable, or unimplemented, automata.
It is an important feature of the methodology presented here that
we can depict operations on systems without depicting the internals
of the systems, thus allowing hierarchical design. The connection
between the operations discussed here and Hoare's parallel operation
is discussed in \cite{KSW:span_graph}, section~4.

For each operation, we describe the effect of the operation on
behaviours, in the sense that we describe the behaviours of the new
system in terms of the behaviours of the given automata. It is an
important feature of our theory that the operations on automata work
fluidly with the notion of behaviour described in
section~\ref{section:behaviour}.

Each operation is described here for two boundary automata --
as noted in section~\ref{section:two_boundary_automata} this is
sufficient to describe it for all automata.

\subsubsection{Binding}
\label{section:bind}

The first operation we consider is binding. Given two
automata with a common boundary, say~$S \colon X \to Y$
and~$T \colon Y \to Z$,
we can produce a new automaton, their \termdef{binding},
denoted~$\bind{S}{T}$. A
state of~$\bind{S}{T}$ is a pair~$(v,w)$, where~$v$ is a state of~$S$
and~$w$ is a state of~$T$.  A motion~$(v,w) \to (v',w')$
of~$\bind{S}{T}$ consists of a pair~$(e,f)$, where~$e \colon v \to v'$
is a motion of~$S$ and~$f \colon w \to w'$ is a motion of~$T$, and
such that~$e$ and~$f$ perform the same action on the boundary~$Y$.
A given motion~$(e,f)$ of~$\bind{S}{T} \colon X \to Z$ is labelled on
the boundary~$X$ by the action~$e$ performs on~$X$ (in~$S$), and is
labelled on the boundary~$Y$ by the action~$f$ performs on~$Z$ (in~$T$).
If each of~$S$ and~$T$ have initial
states~$v_0$ and~$w_0$ respectively, we take~$(v_0,w_0)$ as the initial
state of the binding.

The binding~$\bind{S}{T}$ thus has states the Cartesian
product of the states of~$S$ and~$T$, but motions the subset
of the Cartesian product of motions consisting of those on which
the automata~$S$ and~$T$ synchronize on the common boundary~$Y$. The
reflexive motions of~$T$ allow~$S$ to move independently of~$T$,
provided~$S$ is performing trivial actions on the common boundary.

Binding models two automata communicating by synchronizing on a
common boundary.
We draw diagrams of bound systems by connecting the boundaries of
the automata being bound.

We draw the
binding of two automata~$S \colon X \to Y$ and~$T \colon Y \to Z$
as follows:
\[
\xymatrix
{
X \ar@{-}[r] & \boxedvar{S} \ar@{-}[r] & Y \ar@{-}[r]
& \boxedvar{T} \ar@{-}[r] & Z
}
\]

Importantly, binding interacts well with the behaviours of automata:
\begin{proposition}
Let~$S \colon X \to Y$ and~$T \colon Y \to Z$ be automata with boundary.
To give a behaviour~$\beta$ of~$\bind{S}{T}$ is precisely to
give a behaviour~$\gamma$ of~$S$ and a behaviour~$\delta$ of~$T$ such
that~$\gamma$ and~$\delta$ have the same appearance on~$Y$.
\end{proposition}

\subsubsection{Feedback}
\label{section:feedback}

Given an automaton~$S \colon X \times Y \to Y \times Z$, we can ``bind~$S$
with itself''.  This operation, called \termdef{feedback}, and
denoted~$\fb{Y}{S}$, is used to form closed systems by
connecting boundaries.  We define~$\fb{Y}{S} \colon X \to Z$ to be
the automaton with states precisely those of~$S$, and motions those
motions~$e$ of~$S$ such that~$e$ performs the same action on the factor~$Y$
of its left boundary and the factor~$Y$ of its right boundary.  This
yields an automaton with left boundary~$X$ and right boundary~$Z$, where
the labelling is inherited from~$S$ in the obvious manner. If~$S$
has an initial state~$v_0$, we take~$v_0$ as the initial state of the
fed back automaton.

Just as with binding, feedback may be presented diagrammatically by
connecting the fed back boundaries:
\[
\xymatrix
{
X & \ar@{}[rd]|*{S}&& \\
	&&
	& & Z \ar@{-}[l] \\
&&&
\save "1,3"."2,2"*[F-]\frm{}="frame" \ar@{-}"1,1" \restore
\save "2,2"."1,3" \ar@{-}"2,4"
\ar@{-}`l"3,1"`"3,4"`"1,4"|*{Y}`"frame""frame"
\restore
}
\]

Again, behaviours of the fed back automaton are easy to calculate:

\begin{proposition}
Let~$S \colon X \times Y \to Y \times Z$ be an automaton with boundary.
To give a behaviour~$\beta$ of~$\fb{Y}{S}$ is precisely to
give a behaviour~$\gamma$ of~$S$ such that~$\gamma$ has the same appearance
on the factor~$Y$ of the left boundary as on the factor~$Y$ of the
right boundary.
\end{proposition}

\subsubsection{Product}
\label{section:product}

Given two automata~$S \colon X \to Y$ and~$T \colon Z \to W$, we define
the \termdef{product} of~$S$ and~$T$, an
automaton~$\product{S}{T} \colon X \times Z \to Y \times W$,
in the obvious way -- form the Cartesian product of the states
(resp.~motions) of~$S$ and~$T$ to obtain the states (resp.~motions)
of~$\product{S}{T}$. Note that the
boundaries are likewise formed by Cartesian product -- the product automaton
has boundaries those of~$S$ and those of~$T$. The labelling of
motion~$(e,f)$ is obtained from the labellings of~$e$ and~$f$.
If each of~$S$ and~$T$ have initial
states~$v_0$ and~$w_0$ respectively, we take~$(v_0,w_0)$ as the initial
state of the product.

The product models combining two automata in parallel with no
communication between them. As with binding, we note that
the reflexive actions in the automata allow the automata to act independently.

Diagrammatically, products are shown as follows:
\[
\xymatrix
{
X \ar@{-}[r] & \boxedvar{S} \ar@{-}[r] & Y \\
Z \ar@{-}[r] & \boxedvar{T} \ar@{-}[r] & W
}
\]

Behaviours of the product are easily characterized:

\begin{proposition}
Let~$S \colon X \to Y$ and~$T \colon Z \to W$ be automata with boundary.
To give a behaviour~$\beta$ of~$\product{S}{T}$ is precisely to
give a behaviour~$\gamma$ of~$S$ and a behaviour~$\delta$ of~$T$.
\end{proposition}

\subsubsection{Structural Automata}
\label{section:structural}

\begin{figure}[h]
\[
\xymatrix
{
X && \bullet
		\ar@(ul,dl)[]_{(x|x)}="left"
		\rule[-0.6em]{0em}{1.6em}
	& X
\save
	"left"."1,3"*[F-]\frm{} \ar@{-}"1,1" \ar@{-}"1,4"
\restore
}
\]
\caption{\label{fig:identity} The identity automaton on~$X$}
\end{figure}
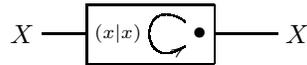

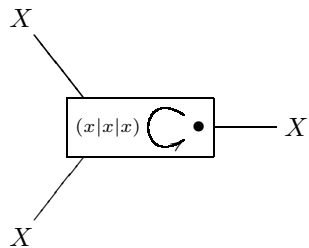
\begin{figure}[h]
\[
\xymatrix
{
X\\
 && \bullet
 		\ar@(ul,dl)[]_{(x|x|x)}="right" 
		\rule[-0.6em]{0em}{1.6em}
	& X \\
X
\save
	"right"."2,3"*[F-]\frm{} \ar@{-}"1,1"
	\ar@{-}"1,1"
	\ar@{-}"3,1"
	\ar@{-}"2,4"
\restore
}
\]
\caption{\label{fig:diagonal} The diagonal automaton on~$X$}
\end{figure}

In addition to the operations on automata, there are a number of constant
operations, or ``structural automata'' which are useful for constructing
systems.  Two examples of note are the identity on a given action set~$X$
(figure~\ref{fig:identity}) and the diagonal on a given action set~$X$
(figure~\ref{fig:diagonal}).

The identity automaton on~$X$ has two boundaries of type~$X$.
It has a single state,
and one motion for each action~$x$ of~$X$,
which is labelled by~$x$ on each boundary.  The reflexive motion is the
motion corresponding to the reflexive action~$\refl \in X$.  As its name suggests,
the identity automaton is the identity for binding on~$X$.  One particular
use of identities is to connect similar boundaries by a single wire
in a composed system.

The diagonal automaton on~$X$ has three boundaries of type~$X$. It
has a single state,
and one motion for each action~$x$ of~$X$,
which is labelled by~$x$ on each boundary.  The reflexive motion is the
motion corresponding to the reflexive action~$\refl \in X$.  The diagonal automaton
is useful for splitting a wire synchronously.

\subsubsection{Binding Philosophers and Forks}

\begin{figure}
\[
\xymatrix
{
&& (0,l) \ar[dd]_{(\al|\refl)}="left"
	&& (0,u) \ar[dd]_{(\al|\refl)} \ar@/^/[rr]^{(\refl|\al)}="top"
		\ar[ddrr]|(.3){(\al|\al)}
	&& (0,r) \ar[dd]^{(\al|\refl)}="right" \ar@/^/[ll]^{(\refl|\au)}
		\ar[ddll]|(.3){(\al|\au)}
	&& \\ &&&&&&&&\\
&& (1,l)
	&& (1,u) \ar@/^/[rr]^{(\refl|\al)} \ar[ddll]^(.3){(\refl|\refl)}
	&& (1,r) \ar@/^/[ll]^{(\refl|\au)}
	&& \\
L &&&&&&&& L\\
&& (2,l) \ar[dd]_{(\au|\refl)}
	&& (2,u) \ar[dd]_{(\au|\refl)} \ar@/^/[rr]^{(\refl|\al)}
		\ar[ddrr]|(.3){(\au|\al)}
	&& (2,r) \ar[dd]^{(\au|\refl)} \ar@/^/[ll]^{(\refl|\au)}
		\ar[ddll]|(.3){(\au|\au)}
	&& \\ &&&&&&&&\\
&& (3,l) \ar`r[ru]`[uuuuuurr]_(.3){(\refl|\refl)}[uuuuuurr]
	&& (3,u) \ar@/^/[rr]^{(\refl|\al)}
	&& (3,r) \ar@/^/[ll]^{(\refl|\au)}="bot"
	&&
\save
	"4,5"."top"."left"."right"."bot"*[F-]\frm{} \ar@{-}"4,1" \ar@{-}"4,9"
\restore
}
\]
\caption{\label{fig:phil_fork} The binding of a philosopher and a fork}
\end{figure}
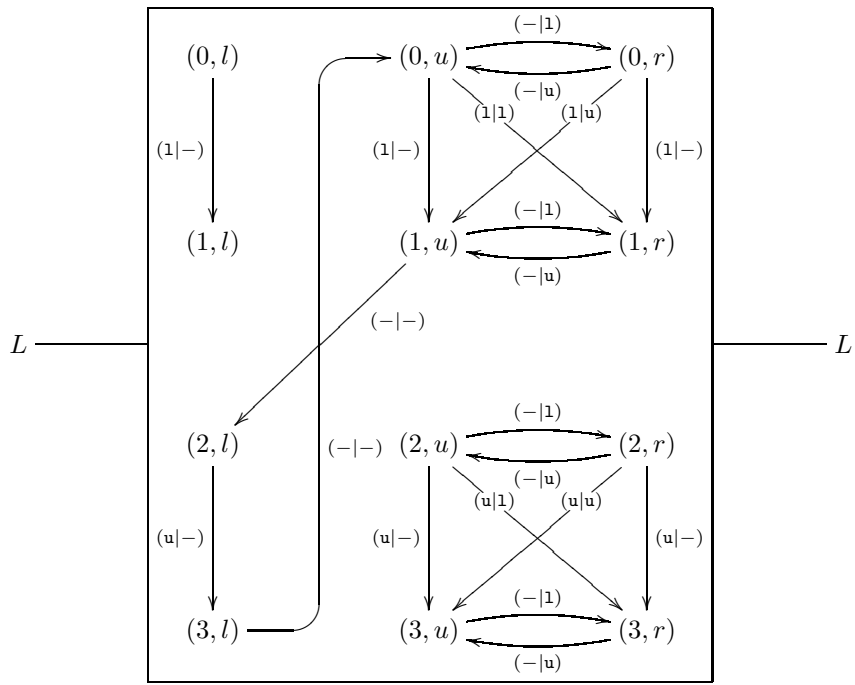

The binding~$\bind{P}{Q}$ of a single philosopher and a single
fork is shown in figure~\ref{fig:phil_fork} -- we have conserved space
a little by abbreviating~\lock\ and~\unlock\ to \al\ and \au\ respectively.
The initial state of the bound system is the state~$(0,u)$.
There are several points to note about the bound system
\begin{itemize}
\item
	Motions where the automata have synchronized have become
	internal motions (i.e., motions that perform trivial
	actions on both boundaries).  In general, synchronizing two motions
	which perform trivial actions on all boundaries that are
	not being synchronized will produce an internal motion.
\item
	Not all states are reachable from the initial state.  Thus
	one often considers the reachable subautomaton of a bound system.
\item
	The model allows for true concurrency,
	not just interleaving semantics. A motion of the bound system such
	as that labelled~$(\al,\al) \colon (0,u) \to (1,r)$ is truly
	concurrent, in that the philosopher and the fork change state
	simultaneously.
\end{itemize}

The binding~$\bind{P}{Q}$ allows a philosopher and a fork to
synchronize on their common boundary by \lock{}ing and \unlock{}ing.

\subsection{Designs and Systems}
\label{section:designs_systems}

The design diagrams we have given for the operations are
more than a guide to the intuition behind the operations.
These diagrams form a precise algebra for constructing designs.
Given a stock of variables for automata with boundaries of given type,
we can draw a diagram by juxtaposing automata and connecting boundaries
with wires for the operations of binding, feedback and product -- such a
diagram is an expression for an automata, which can be evaluated given
automata values for the variables.  Such an expression is called a
\termdef{design}.

\subsubsection{The Geometry of Designs}
\label{section:geometry}

Considering designs as expressions in a precise algebra with the
operations of section~\ref{section:operations}, one
should ``parenthesize'' such expressions to indicate the
desired order of evaluation.

Given two automata~$S \colon X \to Y$
and~$T \colon X \to Y$, we say they are \termdef{isomorphic} if there
is a bijection between states of~$S$ and states of~$T$ and a bijection
between motions of~$S$ and motions of~$T$ which respect the source and
target of motions and the labelling of motions on the boundaries.
One can then prove propositions justifying the diagrammatic manipulations
one would like to carry out, and alleviate the need to parenthesize diagrams
in most situations.

\bigskip

For example, one can easily prove that
binding is associative (up to isomorphism of automata). Given
automata~$S \colon X \to Y$,~$T \colon Y \to Z$ and~$U \colon Z \to W$,
we have that
\[
\xymatrix
{
X \ar@{-}[r]^{}
	& \boxedvar{S} \ar@{-}[r] & Y \ar@{-}[r]
	& \boxedvar{T} \ar@{-}[r]_{} & Z \ar@{-}[r]
	& \boxedvar{U} \ar@{-}[r]
	& W \\
&&& {\cong} &&& \\
X \ar@{-}[r]
	& \boxedvar{S} \ar@{-}[r] & Y \ar@{-}[r]^{}
	& \boxedvar{T} \ar@{-}[r] & Z \ar@{-}[r]
	& \boxedvar{U} \ar@{-}[r]_{}
	& W \\
\save "1,2"."1,4"*!<0.35em,0em>+<1em,1em>\frm{.} \restore
\save "3,4"."3,6"*!<0.35em,0em>+<1em,1em>\frm{.} \restore
}
\]
where the dotted boxes indicate the order of binding.
Symbolically, $\bind{(\bind{S}{T})}{U}$ and $\bind{S}{(\bind{T}{U})}$
are isomorphic.

Thus we can draw diagrams when binding many systems with no risk of
confusion. Of course, binding is not the only operation we consider, and
one can prove propositions relating the different operations:
\begin{proposition}
Given automata with boundary~$S \colon X \to Y$ and~$T \colon Y \to Z$,
we have that
\[
\xymatrix
{
X \ar@{-}[r]^{}
	& \boxedvar{S} \ar@{-}`r[r]`[rdd][dd] & \\
& \boxedvar{T}
	\ar@{-}[r]_{}
	\ar@{-}`l[ld]`[d][d]
	& & Z \ar@{-}[l] & {\cong}
& X \ar@{-}[r] & \boxedvar{S} \ar@{-}[r] & \boxedvar{T} \ar@{-}[r] & Z \\
& Y &
\save "1,2"."2,2".*!<0em,-0.3em>+<1em,1em>\frm{.} \restore
}
\]
In symbols, the automata~$\fb{Y}{\product{S}{T}}$ and~$\bind{S}{T}$ are
isomorphic.
\end{proposition}

The following result is termed the middle four interchange law:

\begin{proposition}
Given automata with boundary~$S \colon X \to Y$, $T \colon Y \to Z$,
$Q \colon U \to V$, and~$R \colon V \to W$, 
we have that
\[
\xymatrix
{
X \ar@{-}[r]^{}
	& \boxedvar{S} \ar@{-}[r] & Y \ar@{-}[r]
	& \boxedvar{T} \ar@{-}[r]^{} & Z
	\\
U \ar@{-}[r]^{}
	& \boxedvar{Q} \ar@{-}[r] & V \ar@{-}[r]
	& \boxedvar{R} \ar@{-}[r]^{}="c12b" & W
	\\
&& {\cong} && \\
X \ar@{-}[r]
	& \boxedvar{S} \ar@{-}[r] & Y \ar@{-}[r]
	& \boxedvar{T} \ar@{-}[r] & Z
	\\
U \ar@{-}[r]
	& \boxedvar{Q} \ar@{-}[r] & V \ar@{-}[r]
	& \boxedvar{R} \ar@{-}[r] & W
\save "1,2"."1,4"*!<0.4em,0em>+<1em,1em>\frm{.} \restore
\save "2,2"."2,4"*!<0.4em,0em>+<1em,1em>\frm{.} \restore
\save "1,2"."2,4"*!<0.7em,-0.5em>+<2em,2em>\frm{.} \restore
\save "4,2"."5,2"*!<0em,-0.3em>+<1em,1em>\frm{.} \restore
\save "4,4"."5,4"*!<0em,-0.3em>+<1em,1em>\frm{.} \restore
\save "4,2"."5,4"*!<0.7em,-0.5em>+<2em,2em>\frm{.} \restore
}
\]
In symbols, the automata~$\product{(\bind{S}{T})}{(\bind{Q}{R})}$
and~$\bind{(\product{S}{Q})}{(\product{T}{R})}$ are isomorphic.
\end{proposition}

It is worth noting that the geometry of designs is a purely combinatorial
geometry -- the wires are a mechanism for denoting the connection between
boundaries, and the curvature and crossing of wires has no effect on
the geometry of the design.
A precise combinatorial model of designs in an appropriate mathematical
context shall be described in a forthcoming paper~\cite{GK:designs}.

\subsubsection{Systems are designs with implementation}
\label{section:system}

By a \termdef{system} we mean a design together with an assignment of
an automaton with boundary to each variable in the design, these assignments
being compatible with the operations used to evaluate designs. An instance
of a variable automaton occurring in a given design is termed a
\termdef{component} of the system. Each
system has an associated automaton, called the \termdef{composite automaton},
or the \termdef{evaluation} of the system, obtained by realising the operations
of the design on the assigned automata in accordance with the definition
of the operations in section~\ref{section:operations}.

While the evaluation of a system may in some sense be seen as the goal
of design, inasmuch as the problem of design is to produce an automaton
with specific properties, the system itself is far more important from
the point of view of analysis.  Retaining the design of the final automaton
in the system allows us to utilise facts about the construction of the
system in order to analyse it -- in section~\ref{section:misa} we shall
give an algorithm which uses design information to assist model checking.
Given the effort typically devoted to design of a system in real world terms,
it seems only rational that a theory of distributed systems retain designs
as an element which is both precisely represented and capable of being
computed with. For even though
the ultimate (external customer) deliverable of a development effort is
the compiled code (= evaluated automaton), a development effort is also
expected to deliver a design to maintenance engineers in a
usable (= analyzable) form.

\subsubsection{Subsystems}
\label{section:subsystem}

Given a system, by a \termdef{subsystem} we mean some subset of the components
of the system.  Such a subsystem gives rise to an automaton via evaluation,
we shall usually abuse terminology and use the term subsystem for this
evaluation also.

For both binding and product, the states of the constructed automaton are
pairs of states of the automata being operated upon.  In the case of feedback,
the states are states of the fed back automaton. Hence each state of the
evaluation of a system gives rise to a state of the automaton associated with
each component.  We shall refer to states of the automata associated with
components as \termdef{local states}, and by contrast refer to a state~$v$
of the evaluation of the system as a \termdef{global state} of the system.

Further, given a subsystem of
some larger system, each state of the evaluation of the system gives rise
to a state of the subsystem in the obvious way (subsystem states are
tuples of the local states of the components which form the subsystem).
We refer to a state of a subsystem arising in this manner as a
\termdef{local state} of the subsystem.

Similar remarks hold for motions and behaviours, and we shall thus use
the terms \termdef{local motion}, \termdef{global motion},
\termdef{local behaviour}, and \termdef{global behaviour} for the
corresponding concepts.

\subsubsection{Dining Philosopher Systems}

\begin{figure}[h]
\[
\xymatrix
{
&
\boxedvar{P} \ar@{-}[r] & \boxedvar{Q} \ar@{-}[r] &
\boxedvar{P} \ar@{-}[r] & \boxedvar{Q} \ar@{-}[r] &
\boxedvar{P} \ar@{-}[r] & \boxedvar{Q}
	\ar@{-}`r[r]`[d]`[dllllll]`[llllll][lllll] & \\
&&&&&&&
}
\]
\caption{\label{fig:phil_3_ring} A ring of 3 philosophers and their forks}
\end{figure}
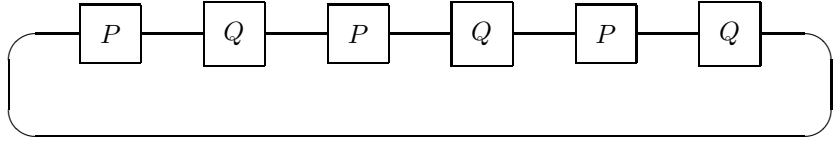

For any~$n \in \N$, we can form the composite
automaton~$\fb{L}{(\bind{P}{Q})^{n}}$ - this automaton has no boundaries,
and models a ring of~$n$ philosophers with their~$n$ intervening forks.
For example, figure~\ref{fig:phil_3_ring} shows a design for a ring of
three philosophers with their forks. This design, together with with
assignment of~$P$ and~$Q$ to the automata of figures~\ref{fig:phil}
and~\ref{fig:fork} respectively, comprise a system of dining philosophers.

\subsection{Linear Automata}
\label{section:linear}

A motion~$e$ of an automaton with boundary~$(S,(X_i,\mu_i)_{i \in I})$
is said to be \termdef{linear} if the action~$\mu_i(e)$ performed
by~$e$ on the~$i$'th boundary is nontrivial for at most one~$i \in I$.
The automaton~$S$ itself is said to be \termdef{linear}
if every motion of~$S$ is linear.

That is, a linear automaton is an automaton that interacts with at most one
boundary in a given state.  Linear automata have their boundaries decoupled,
in the sense that they never require simultaneity on distinct boundaries.
Another point of view is that linear automata are those modelling
systems for which {\em interleaving} semantics are sufficient.

We note that even if the automata~$S$ and~$T$ are linear,
the binding~$\bind{S}{T}$ and the product~$\product{S}{T}$ may
be nonlinear. For example, the binding of a philosopher and a fork
(each a linear automaton) produces the nonlinear automaton of
figure~\ref{fig:phil_fork}.

\subsubsection{Linearizable Automata}
\label{section:linearizable}

An automaton with boundary~$(S,(X_i,\mu_i)_{i \in I})$ is
\termdef{linearizable} if, for each motion~$e \colon v \to w$ of~$S$
and given total order on~$I$, we can find a behaviour
\[
\begin{diagram}
v = v_0 & \rTo^{e_1} & v_1 & \rTo^{e_2} & \ldots & \rTo^{e_n} & v_n = w
\end{diagram}
\]
of~$S$ such that
\begin{enumerate}[(i)]
\item each~$e_k$ is linear
\item\label{resp_order}
	if~$k, l \in [n]$ are such that~$\mu_i(e_k)$
	and~$\mu_j(e_l)$ are nontrivial and~$i \leq j$ in the total order on~$I$,
	then we have ~$k \leq l$
\item\label{complete}
	if~$i \in I$ is such that~$\mu_i(e)$ is nontrivial, then
	there exists a~$k \in [n]$ such that~$\mu_i(e) = \mu_i(e_k)$.
\end{enumerate}
where~$[n]$ denotes the set~$\{ 1, \ldots, n\}$.
Note that condition~(\ref{resp_order}) implies that distinct~$e_k$'s cannot
both perform nontrivial actions on the same boundary, i.e. that the existence
in~(\ref{complete}) is unique.

Less symbolically, linearizable
automata are those for which any nonlinear motion~$e$ can be refined into
a series of linear motions with any desired ordering on the actions
carried out simultaneously by~$e$.

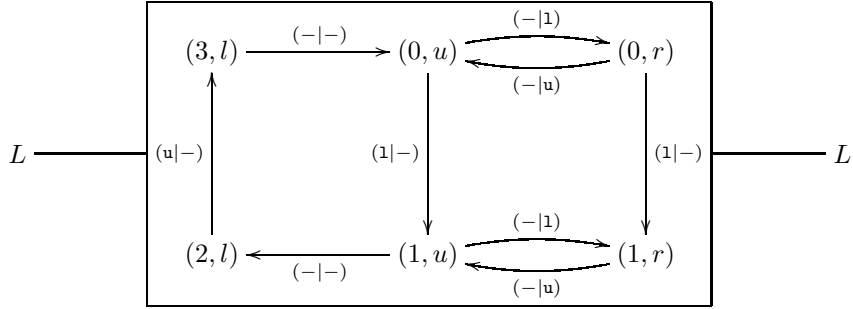
\begin{figure}[h]
\[
\xymatrix
{
&& (3,l) \ar[rr]^{(\refl|\refl)}
	&& (0,u) \ar[dd]_{(\al|\refl)} \ar@/^/[rr]^{(\refl|\al)}="top"
	&& (0,r) \ar[dd]^{(\al|\refl)}="right" \ar@/^/[ll]^{(\refl|\au)}
	&& \\ L&&&&&&&&L\\
&& (2,l) \ar[uu]^{(\au|\refl)}="left" 
	&& (1,u) \ar@/^/[rr]^{(\refl|\al)} \ar[ll]^{(\refl|\refl)}
	&& (1,r) \ar@/^/[ll]^{(\refl|\au)}="bot"
	&&
\save
	"2,5"."top"."left"."right"."bot"*[F-]\frm{} \ar@{-}"2,1" \ar@{-}"2,9"
\restore
}
\]
\caption{\label{fig:phil_fork_reachable}
	The linearized reachable binding of a philosopher and a fork}
\end{figure}

Given linear automata~$S$ and~$T$, we observed above their binding
and product need not be linear.  It is however the case that they
will be linearizable.
A linearizable automaton can be linearized by considering only
the linear motions. For example, linearizing the
subautomaton of figure~\ref{fig:phil_fork} reachable from
the initial state produces the automaton shown in
figure~\ref{fig:phil_fork_reachable}.

It should be noted that one cannot restrict attention solely to linear
automata, as the operation of feedback presented in
section~\ref{section:feedback} is not well suited to linear automata --
the only motions in a fed back linear automaton are those which are trivial on
the fed back boundaries.  Further, the structural components described in
section~\ref{section:structural} are not linear automata.

\subsubsection{Atomic Motions}
\label{section:atomic}

Consider a system in which each automaton assigned to variable of the design
is linear.  Given a global motion~$e$ of the system, we have a corresponding
local motions~$e_c$ for each component~$c$ of the system.  We say that
the global motion~$e$ is an \termdef{atomic motion} if
\begin{enumerate}[(i)]
\item The local motions~$e_c$ are nontrivial for at most two components.
\item In the case that the local
	motions~$e_c$ and~$e_d$ are nontrivial for distinct
	components~$c$ and~$d$, then the components~$c$ and~$d$ have
	boundaries~$(X,\mu_i)$ and~$(X,\mu_j)$ (respectively) which are joined
	by a wire in the design, and for which~$\mu_i(e_c) = \mu_j(e_d)$ is
	nontrivial.
\end{enumerate}
That is to say, the atomic motions are those for which multiple components
move nontrivially only in the event they are synchronizing on boundaries
joined by the design.

Given a system comprised of linear automata, the subautomaton of the
composite automaton with all states but only atomic motions is termed the
\termdef{atomic core} of the system.  The atomic core restricts attention
to those motions which are not fortuitously simultaneous.  We shall use this
notion when discussing model checking for deadlock in
section~\ref{section:misa}.  We simply remark at this point that when
considering a complete system, the atomic core allows sufficient motions to
fully explore the system, in the sense that any state~$w$ of the system
reachable from a state~$v$ is reachable via atomic motions.
However, we cannot restrict attention to
the atomic core prematurely, for feedback of systems relies on nonlocal
simultaneity.

\bigskip

We shall not investigate linearity further at present, but merely note
that one of the strengths of the theory presented here is that specific
requirements for envisaged domains (e.g. interleaving semantics) can be
carried as additional properties of, or structure on, the basic theory.
Precisely what can be done to tailor the basic theory for application to
a specific domain is an area for further interesting work. Further work
on linear automata can be found in section~4.1 of~\cite{KSW:span_graph},
including a connection with the process algebras of Hoare.

%% file: model.tex
In this section we turn our attention to the problem of {\em model
checking} -- verifying that a given system has certain properties.
The property we shall examine in detail is that of deadlock. We give
an algorithm for finding a subspace of the state space of a given automaton,
such that if the automaton possesses a deadlock~$v$, then the subspace
possesses~$v$.
In the case of the dining philosophers, this subspace is only quadratically
large in the number of philosophers.  This result has also been achieved
using stubborn sets (see~\cite{valmari:state_art}).
We then indicate some examples where the algorithm does not give such
a good result. In section~\ref{section:sim_checking}, we shall show
how to leverage the algorithm presented here to these cases using
abstraction techniques.

For the remainder of this section, we restrict attention to finite
automata.

\subsection{Deadlock detection}
\label{section:deadlock_detection}

A state~$v$ of a automaton~$S \colon X \to Y$ is said to be a
\termdef{deadlock}
state  if the only motion with source~$v$ is the reflexive motion.
For example, if the composite of three dining philosophers and their
forks (figure~\ref{fig:phil_3_ring}) is evaluated, the
state~$(1,r,1,r,1,r)$ is a reachable deadlock.
Naively, to check for a reachable deadlock in an automaton, one must examine
every reachable state and determine if it is a deadlock state. In the
example of the dining philosophers, a ring of~$n$ philosophers
has~$(4 \times 3)^{n}$ states, of which $3^n - 1$ are reachable
(for $n \geq 2$).

However, we can attempt to exploit the design of the system to simplify
our search.  The motivating case is the example of products of automata:
\begin{proposition}
\label{prop:product_deadlock}
Let~$S \colon X \to Y$ and~$T \colon Z \to W$ be automata with given
initial states~$v_0$ and~$w_0$.  If the product~$S \times T$ has a reachable
deadlock, then it has a reachable deadlock of the form~$(v^\ast,w^\ast)$
where~$v^\ast$ is a deadlock of~$S$ and~$w^\ast$ is a deadlock of~$T$.

Moreover, this deadlock is reachable by first considering those motions
trivial in~$T$, and then considering those motions trivial in~$S$.
\end{proposition}

The content of the proposition is that it suffices to check the
subautomaton of states of the form~$(v,w_0)$ or~$(v^\ast,w)$ when searching
for a reachable deadlock.  In general, this subautomaton has a number of
states bounded by~$\#S + \#T$, as opposed to the~$\#S \times \#T$ states
in the full product~$S \times T$.

We define a \termdef{strong deadlock analysis} of an
automaton~$S \colon X \to Y$ with initial state~$v$ to be a
subautomaton~$T$ of~$S$ such that
\begin{enumerate}[(i)]
\item $T$ contains~$v$
\item If~$w$ is a deadlock state of~$S$ reachable from~$v$,
	then~$w$ is in~$T$.
\end{enumerate}
One could also consider the notion of a \termdef{weak deadlock
analysis}, where the second condition is replaced by the
weaker condition
\begin{enumerate}
\item[(ii')] If a deadlock state of~$S$ is reachable from~$v$,
	then~$T$ contains a deadlock reachable from~$v$.
\end{enumerate}

Then the content of the proposition~\ref{prop:product_deadlock} is that
the subautomaton of~$S \times T$ with states those states~$(v,w)$ such
that~$v = v^\ast$ or~$w = w_0$, and motions all motions between these
states, comprises a weak deadlock analysis of the product.  If we include
all states~$(v,w)$ such that either~$v$ is a deadlock of~$S$ or~$w = w_0$ we
obtain a strong deadlock analysis of the product.

\subsection{Introspective Subsystems}
\label{section:introspective}

Most systems of interest do not decompose as products as required for the
deadlock analysis provided by proposition~\ref{prop:product_deadlock} --
some coupling is required in order for distributed parts of the system
to communicate and achieve a common goal. However, many systems of
interest do ``locally decouple'' in the sense that parts do not spend
their entire time in communication with each other, and generally restrict
their interaction to specific parts of the system at specific times.

Before presenting a notion of local decoupling appropriate to our theory,
we mention again the example of the dining philosophers. Consider the
reachable linearized subautomaton of the binding of a philosopher
and a fork as shown in figure~\ref{fig:phil_fork_reachable}. From the
state~$(3,l)$ we have a motion to~$(0,u)$ labelled~$(\refl|\refl)$.  Being
an internal motion of the bound automaton and the only motion out of~$(3,l)$,
any behaviour from this state must use this motion, and the environment of
the automaton cannot affect the viability of this motion.  That is to say,
the philosopher and the fork have locally decoupled from the rest of any
larger system they may be a part of.  We would like to restrict our search
for a deadlock by taking advantage of the fact this motion is independent
of the action of the rest of the system.

Given an automaton~$S$ with boundaries~$(X_i,\mu_i)$, and a state~$v$ of~$S$,
we say that~$S$ is \termdef{watching} or \termdef{looking at}
boundary~$i$ in state~$v$ if there is a motion~$e \colon v \to w$
of~$S$ such that~$\mu_i(e)$ is nontrivial.
Conversely, the automaton~$S$ is \termdef{ignoring} boundary~$i$
in state~$v$ if every motion with source~$v$ performs the trivial
action on the boundary~$X_i$.

For example, the philosopher of figure~\ref{fig:phil} is watching the left
boundary in states~$0$ and~$2$, and watching the right boundary in states~$1$
and~$3$.  The fork of figure~\ref{fig:fork} is watching the left boundary
in states~$l$ and~$u$, and the right boundary in states~$u$ and~$r$.

\begin{proposition}
\label{prop:can_move_S}
Let~$S \colon X \to Y$ and~$T \colon Y \to Z$ be automata. Let~$v$
be a nondeadlock state of~$S$ such that~$S$ is ignoring~$Y$ in
state~$v$. For any state~$w$ of~$T$, if there is a
behaviour of~$\bind{S}{T}$ with initial state~$(v,w)$ which
leads a deadlock~$(v^\ast, w^\ast)$, then
there is such a behaviour where the first nontrivial motion is
trivial in~$T$.
\end{proposition}
\begin{proof}
Suppose, by way of contradiction, this were not true.  Let~$(v,w)$
be a state of~$\bind{S}{T}$ such that a deadlock~$(v^\ast, w^\ast)$
is reachable from~$(v,w)$,
but not by an initial nontrivial motion which is trivial in~$T$.
Let~$\beta$ be a behaviour
of~$\bind{S}{T}$ with initial state~$(v,w)$ and reaching the deadlock.
Write
\[
\begin{diagram}
\beta = (v_0,w_0) & \rTo^{(f_1,g_1)} &
	(v_1,w_1) & \rTo^{(f_2,g_2)} &
	(v_2,w_2) & \ldots & (v_n,w_n) = (v^\ast, w^\ast)
\end{diagram}
\]
where~$v_0 = v$ and~$w_0 = w$.

We claim some~$f_k$ is a nontrivial motion of~$S$.  If not,
then~$v_n = v$. Since~$v$ is not a deadlock state of~$S$, there is
some motion with source~$v$, say~$e$ labelled~$(x|y)$.  Since~$S$ is not looking
at~$Y$ in state~$v$, it must be that~$y = \refl$ is trivial.  Hence we
can extend~$\beta$ with the motion~$e$ in~$S$ and the trivial motion
in~$T$, and thus~$\beta$ did not reach a deadlock state, contrary to choice
of~$\beta$.

Let~$k$ be minimal such that~$f_k$ is a nontrivial motion in~$S$.
Note that~$f_i$ is trivial for~$i = 0$,~$\ldots$,~$k-1$.  Thus
\[
v_0 = v_1 = \ldots = v_{k-1}
\]
The triviality of~$f_i$ for~$i < k$ also implies that the action
performed by~$f_i$ on the boundary~$Y$ is trivial.  The action performed
by~$f_k$ on the boundary~$Y$ is also trivial, since~$S$ is not looking
at~$Y$ in state~$v$.  Thus, since~$g_i$ synchronizes with~$f_i$, we
have that the action performed by~$g_i$ on the boundary~$Y$ is also
trivial, for~$i = 0$,~$\ldots$,~$k$.

It is now evident that
\[
\begin{diagram}
(v_0,w_0) & = (v_{k-1},w_0) & \rTo^{(f_k,\refl)} &
	(v_k,w_0) & \rTo^{(\refl,g_1)} &
	(v_k,w_1) \\
\quad & \rTo^{(\refl,g_2)} (v_k,w_2) & \ldots
	(v_k,w_{k-1}) & \rTo^{(\refl,g_k)} &
	(v_k,w_k) \\
\quad & \rTo^{(f_{k+1},g_{k+1})} &
	(v_{k+1},w_{k+1}) & \ldots & (v_n,w_n) = (v^\ast, w^\ast)
\end{diagram}
\]
is a behaviour of~$\bind{S}{T}$ with initial state~$(v,w)$
that leads to the specified deadlock~$(v^\ast, w^\ast)$,
and that has first motion trivial in~$T$. Hence the desired contradiction.
\end{proof}

For a system with composite automata~$S$ and a global state~$v$ of~$S$,
a subsystem is said to be \termdef{introspective at~$v$} if each component
of the subsystem, when in the local state corresponding to~$v$, is
ignoring every boundary on which it connects to components not in
the subsystem.

\begin{proposition}
\label{prop:introspect}
Consider a global state~$v$ of a system and a subsystem that is
introspective at~$v$ but not deadlocked when in the local state
corresponding to~$v$. If a deadlock of the system is reachable from~$v$,
it is reachable via a behaviour whose first nontrivial motion is
trivial outside the given subsystem.
\end{proposition}
\begin{proof}
Using the algebra of designs, organize the system as a composite
of the given subsystem and its complement:
\[
\xymatrix
{
& *+<1.5em>[F-]\txt{Introspective\\Subsystem}
	\ar@{-}[l]
	\ar@{-}@<1ex>[r]
	\ar@{-}[r]
	\ar@{-}@<-1ex>[r]
& *+<1.5em>[F-]\txt{Rest of\\System}
	\ar@{-}[r] &
}
\]
Now apply proposition~\ref{prop:can_move_S} to the composite of the evaluation
of the two subsystems.
\end{proof}

Given an introspective but not deadlocked subsystem at each global
state~$v$ of the composite automata~$S$ of a system, we can apply
proposition~\ref{prop:introspect} repeatedly to produce a strong
deadlock analysis by including only those motions of~$S$ which are trivial
outside the introspective subsystem associated with their source,
and including only those states which are reachable from
the initial state of~$S$ via the included motions.

\subsection{Minimal Introspective Subsystem Analysis}
\label{section:misa}

It may be that, for a given design, the introspective subsystems are
obvious, or designed in to the system so as to provide for
more efficient checking. However, it is also desirable to automatically check
a given system for absence of deadlock, exploiting the known design
of the system to reduce the state space explosion associated with
exhaustive model checking.

The idea of minimal introspective subsystem analysis is to guide the exploration
of the state space via proposition~\ref{prop:introspect}.
More precisely, we construct the deadlock analysis
of a given system suggested at the end of the the previous section
as we explore the state space, by choosing a minimal introspective
subsystem at each state.

Let us fix for discussion a system with composite automaton~$S$.
Given a global state~$v$ of~$S$ and a component of the system,
we can examine the automaton assigned to the component to determine
which boundaries the automaton is looking at in the local state corresponding
to~$v$.

Given this information for each component and the design of the
system,  it is a simple matter to
construct a non-deadlocked minimal introspective subsystem at~$v$
-- consider the directed graph with vertices the components and edges
indicating that the component represented by the source is looking at
the component represented by the target, and flood fill along edges from
each vertex to find introspective subsystems.

This process determines, for each component, the minimal introspective
subsystem containing that component.  If the global state~$v$ is not
a (global) deadlock, then some nontrivial motion is possible.  Hence at
least one component, and thus the introspective subsystem containing it,
is not deadlocked.  We select the smallest of these
subsystems which is not deadlocked in the local state corresponding to~$v$.

We then explore the states of~$S$ only along motions
which are trivial outside
the selected minimal introspective subsystem, looking for deadlock --
proposition~\ref{prop:introspect} guarantees that a reachable deadlock
is reachable via a motion trivial outside the introspective subsystem.
By maintaining a list of visited states, we can ensure the algorithm
terminates.

In the event the automata assigned to variables of the design are linear,
it suffices to explore the states of~$S$ only along atomic motions
(see~\ref{section:atomic}) -- in this case restricting to the subautomaton
including only atomic motions does not alter the reachability of states.

\subsubsection{Minimal Introspective Subsystem Analysis of Dining Philosophers}
\label{section:misa_dp}

Let us consider the system consisting of a ring of~$n$ philosophers
and their~$n$ intervening forks. Since each automata used in the system
is linear, we can restrict our attention to atomic motions.
We shall now walk through the application of the minimal introspective
subsystem analysis algorithm proposed in section~\ref{section:misa} for
this system.

Initially, each philosopher is looking at the left boundary, and each
fork is looking at both boundaries.  So the only introspective
subsystem is the entire system, giving~$n$ atomic motions to be explored
(each motion being one for which a given philosopher acquires their left
fork).

Each state reached next has precisely one philosopher having obtained their
left fork. Let us consider the state in which the philosopher~$P_1$
has acquired his left fork.  At this point, we see~$P_1$ is now
looking at his right boundary, and~$P_2$ is looking at her left boundary,
and the fork~$F_1$ is looking at both boundaries.  These three components
thus form an introspective subsystem as required.  A moments thought
shows that it is minimal, and a moments more that it is the only minimal
introspective subsystem.  There are only two nontrivial atomic
motions in this subsystem -- either~$P_1$ acquires the fork or~$P_2$
acquires the fork.

In the former case, the system comprised of~$P_1$ and his left fork now
constitute a minimal introspective subsystem -- the philosopher is
in state~$2$ attempting to relinquish the left fork, and the left
fork is in state~$r$ having being acquired by the philosopher on its
right.  The single nontrivial atomic motion of the subsystem is to
relinquish the fork.  Following this, and by a similar
analysis, the philosopher relinquishes his right fork.  The system has
now returned to the initial state, which is marked as checked.

In the latter case of~$P_2$ acquiring the fork~$F_1$, we apply similar
reasoning to the competition between~$P_2$ and~$P_3$ to acquire~$F_2$,
as these three systems once again comprise the unique minimal
introspective subsystem at the global state under consideration.
In exploring the case that~$P_2$ is successful, she will
proceed to relinquish her left and right forks, and we return to the state
where she is competing with~$P_1$.  In exploring the case~$P_3$ is successful
we examine the subsystem comprised of~$P_3$,~$F_3$ and~$P_4$.

The algorithm continues in this manner, obtaining a minimal introspective
subsystem comprising two philosophers and their intervening fork at
each stage, and progressing in two ways -- allowing one philosopher to run
to completion, or moving to the competition for the next fork around the
table.  Eventually the deadlock in which each philosopher has acquired
their left fork is found.

It is evident that after the initial state, we explore 3 states for each
philosopher (as it moves through states~$1$,~$2$ and~$3$) bar the last.
The exploration stops when the last philosopher acquires his left fork,
and hence each philosopher has acquired their left fork and the system
is deadlocked.  Potentially then, we are required to explore the initial
state, the final deadlock state, and~$3(n-1)$ for each choice of the
initial~$n$ motions.  Thus~$3n^2 - 3n + 2$ states are explored, a
significant reduction on the~$3^n-1$ reachable states in the system.

This result has been described using stubborn sets
in~\cite{valmari:error_detect}, where the same polynomial for the number
of states checked is computed.

\medskip

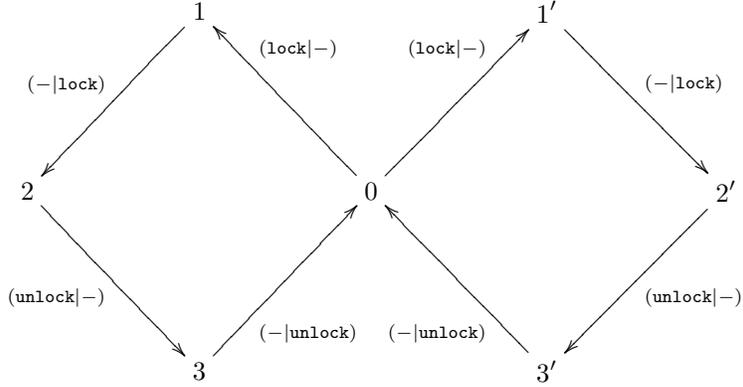
\begin{figure}[th]
\[
\xymatrix
{
&& 1 \ar[lldd]_{(\refl|\lock)}
	&&&&
	1' \ar[rrdd]^{(\refl|\lock)} &&
	\\ \\
2 \ar[rrdd]_{(\unlock|\refl)}="left"
	&&&& 0 \ar[lluu]_(0.7){(\lock|\refl)} \ar[rruu]^(0.7){(\lock|\refl)}
	&&&& 2' \ar[lldd]^{(\unlock|\refl)}="right"
	\\ \\
&& 3  \ar[rruu]_(0.3){(\refl|\unlock)}
	&&&& 3'  \ar[lluu]^(0.3){(\refl|\unlock)}
	&&
}
\]
\caption{\label{fig:nondet_phil} Alternative Philosopher I}
\end{figure}

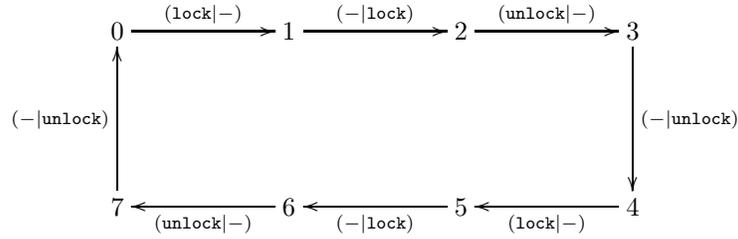
\begin{figure}[th]
\[
\xymatrix
{
0 \ar[rr]^{(\lock|\refl)}
	&& 1 \ar[rr]^{(\refl|\lock)}
	&& 2 \ar[rr]^{(\unlock|\refl)}
	&& 3 \ar[dd]^{(\refl|\unlock)}
	\\ \\
7 \ar[uu]^{(\refl|\unlock)}
	&& 6 \ar[ll]^{(\unlock|\refl)}
	&& 5 \ar[ll]^{(\refl|\lock)}
	&& 4 \ar[ll]^{(\lock|\refl)}
}
\]
\caption{\label{fig:slow_phil} Alternative Philosopher II}
\end{figure}

It should be noted that there are systems very similar to the dining
philosophers in which the above algorithm does not reduce the checking
to a polynomial number of states.  Replacing the philosophers by either
the system shown in figure~\ref{fig:nondet_phil} or
figure~\ref{fig:slow_phil} results in a system for which the above algorithm
searches an exponential number of states.

The alternative philosopher I, shown in figure~\ref{fig:nondet_phil},
may be termed the nondeterministic philosopher.  In this case,
minimal introspective subsystem analysis of the composite system must check
two branches as the minimal introspective subsystem under consideration
moves around the table.  Informally then, we see that an exponential
number of states will be checked (although still significantly less than
the total number of states of the system).

The alternative philosopher II, shown in figure~\ref{fig:slow_phil},
may be termed the double cover philosopher.  In this case,
when the philosopher returns to the state of wishing to acquire his
left fork in the first instance, he is in the local state~4, and not
a searched state as in the basic example. Thus the algorithm 
arrives at many distinct states in which the minimal introspective
subsystem is the entire system.  There are~$2^n$ reachable
global states of this form, and it can be argued that the algorithm
proposed above will visit all of them.  Informally then we again
have a situation where an exponential number of states are checked.

It is worth noting that both the nondeterministic and double cover
philosophers can be abstracted to the philosopher of figure~\ref{fig:phil}
in a sense which is made precise in section~\ref{section:simulations}.
This abstraction allows us to use the strong deadlock analysis
with only polynomially many states to check the more complex systems
(see section~\ref{section:sim_checking}).

\bigskip

In this section we have outlined the principles of model checking for
deadlock as manifested in our theory, and presented a very simple algorithm
for reducing state space explosion in model checking.  It should be
emphasized that although simplistic, the algorithm does have demonstrably
good behaviour on a particular system, and importantly exhibits the principle
of using design information retained by our calculus to assist in the
process of model checking.

%% file: sim.tex
In this section we introduce a compositional notion
of simulation of automata which has a close relation to
work on simulations and bisimulations
(see~\cite{abramsky:interaction_categories},~\cite{CS:sync_async},
and~\cite{JNW:bisim_open}).
We will also indicate how simulations can be
used to facilitate model checking.

\subsection{Comparison of Automata}
\label{section:comparison}

We begin by defining reflexive graph morphisms. Denote a 
reflexive graph~$G$ by the pair~$(V,E)$ comprising its set of
vertices~$V$ and its set of edges~$E$. A
\termdef{reflexive graph morphism}~$f$ from~$G=(V,E)$ to~$G'=(V',E')$
consists of functions~$f_V\colon V \to V'$ and~$f_E\colon E \to E'$
such that sources and targets of edges are preserved,
as are reflexive edges.

Suppose~$(S,(X_i, \mu_i)_{i \in I})$ and~$(T,(X_i, \nu_i)_{i \in I})$
are two automata with the same boundary action sets~$(X_i)_{i \in I}$.
A \termdef{comparison}~$f$ from~$S$ to~$T$ is a reflexive graph
morphism~$f$ from~$S$ to~$T$ which preserves
the actions on the boundaries -- that is, for each~$i \in I$,
we have~$\nu_i \cdot f = \mu_i$.  We shall term the function comprising the
action of~$f$ on states the \termdef{state map}, and the function comprising
the action of~$f$ on motions the \termdef{motion map}.
 
When we are writing automata in the form~$X \to Y$ 
(as two boundary automata),
such a comparison is denoted~$f\colon S \compto T\colon X \to Y$,
although typically we shall just write~$f\colon S \compto T$ as
the boundaries will be
understood.  In this section we will consider automata
equipped with an initial state, and comparisons are asked to
preserve initial states.

For example, let~$P\colon L \to L$ be the philosopher described in
section~\ref{section:dp_automata} (figure~\ref{fig:phil}),
let~$P'\colon L \to L$ be the alternative philosopher I depicted
in figure \ref{fig:nondet_phil} and let~$P''\colon L \to L$ be the
alternative philosopher II depicted in figure~\ref{fig:slow_phil}.
There are unique comparisons~$p\colon P' \compto P$,~$q\colon P'' \compto P$
and~$r\colon P'' \compto P'$, and  these comparisons preserve the
initial vertex~$0$.
 
\medskip
 
Action sets, automata and comparisons form what is known as 
a {\em discrete Cartesian bicategory} (see~\cite{CW:cart_bicat}).
Rather than recalling the definition
of this complicated algebraic structure, we will only consider 
the operations relevant to this paper; namely,
composition, binding, feedback and product of comparisons.

\subsubsection{Composition}
\label{section:comparison_composition}

Given automata~$R$,~$S$, and~$T \colon X \to Y$, and
comparisons~$f \colon R \compto S$ and~$g \colon S \compto T$,
we define the \termdef{composite}
comparison~$\compose{g}{f}\colon R \compto T$. The composite has
state map the composite of the state maps of~$f$ and $g$ and
motion map the composite of the motion maps of~$f$ and~$g$ -- both
these latter composites being the usual composite of functions.
It is routine to check that the composite, so defined, is
a comparison~$R \compto T$.

In categorical terms, this composite is simply the composite of~$f$
and~$g$ in the category of reflexive graphs.
In fact, for any two action sets~$X$~and~$Y$ we
can form a category~${\bf Aut}(X,Y)$. Its objects are
automata of the form~$S\colon X \to Y$ and its arrows are
comparisons between these automata.

\subsubsection{Binding, feedback and product}
\label{section:comparison_bfp}

We now describe how the operations on automata described in
section~\ref{section:operations} also apply to comparisons.  In each
case, the data for the operations consists of comparisons between
data suitable for the corresponding automata operation.

Given automata~$S$ and~$T \colon X \to Y$ and
automata~$U$ and~$V \colon Y \to Z$,
together with
comparisons~$f \colon S \compto T$ and~$g \colon U \compto V$, 
we define the \termdef{binding} of~$f$~and~$g$ -- a
comparison~$\bind{f}{g}\colon \bind{S}{U} \compto \bind{T}{V}$.

As described in section~\ref{section:bind}, the state space
of~$\bind{S}{U}$ has states consisting of pairs~$(v,w)$ with~$v$
a state of~$S$ and~$w$ a state of~$U$, and motions consisting of pairs
of motions performing the same action on the common boundary.  The
comparison~$\bind{f}{g}$ maps the state~$(v,w)$ by mapping~$v$ as~$f$
does, and~$w$ as~$g$ does, with the obvious extension to motions.

The fact that~$f$ and~$g$ respect the actions of motions on boundaries
implies that~$\bind{f}{g}$ maps motions of~$\bind{S}{U}$ to motions
of~$\bind{T}{V}$, and it is routine to check we have defined a comparison.

\medskip

Given automata~$S$ and~$T \colon X \times Y \to Y \times Z$
and a comparison~$f \colon S \compto T$,
we define the \termdef{feedback} of~$f$ -- a
comparison~$\fb{Y}{f} \colon \fb{Y}{S} \compto \fb{Y}{T}$.

The state space of~$\fb{Y}{S}$ is that of~$S$, but with motions only
those that perform the same action on the two boundaries of~$S$ of type~$Y$.
Since~$f$ preserves the actions on boundaries, the image of such a motion
under~$f$ is a motion of~$T$ performing the same action on the two boundaries
of~$T$ of type~$Y$.  Thus the comparison~$f$ restricts to a comparison
between the fed back automata, and this latter comparison is~$\fb{Y}{f}$.

\medskip

Given automata~$S$ and~$T \colon X \to Y$ and
automata~$U$ and~$V \colon W \to Z$, together with
comparisons~$f \colon S \compto T$ and~$g \colon U \compto V$, 
we define the \termdef{product} of~$f$~and~$g$ -- a
comparison~$\product{f}{g}\colon \product{S}{U} \compto \product{T}{V}$.

The product~$\product{f}{g}$ has state (resp. motion) map the product of the
state (resp. motion) maps of~$f$ and~$g$.  That is, it operates on the
state space of~$\product{S}{U}$ componentwise.

\subsection{Simulations}
\label{section:simulations}

If~$S\colon X \to Y$ is an automaton, let~$\reachable{S}\colon X \to Y$
denote the reachable subautomaton of~$S$. Note that to give a
comparison~$f\colon\reachable{S} \compto T$ is just to give a
comparison~$f\colon\reachable{S} \compto \reachable{T}$.

By a \termdef{simulation}~$f$ from~$S\colon X \to Y$
to~$T\colon X \to Y$ we mean a
comparison~$f\colon \reachable{S} \compto \reachable{T}$
such that~$f$ satisfies the
following `lifting property': for all states~$v$ of $\reachable{S}$
and all motions~$e\colon f(v) \to w$ in $\reachable{T}$,
there exists a (finite) behaviour of $\reachable{S}$
\[
\begin{diagram}
v=v_0 & \rTo^{e_0} & v_1 & \rTo^{e_1} & \ldots & \rTo^{e_n} & v_{n+1} 
\end{diagram}
\]
such that
\begin{enumerate}[(i)]
\item
	for~$0 \leq i \leq n-1$, the motion~$f(e_i)$ is the reflexive
	motion at~$f(v)$
\item
	$f(e_n) = e$
\end{enumerate}
We say that the automaton~$T$
\termdef{simulates~$S$ via~$f$}, and
write~$f\colon S \simto T\colon X \to Y$,
or merely  $f\colon S \simto T$ if the boundaries are understood.

\medskip

In light of this definition, we should revisit our view of reflexive
motions as idling motions.  More precisely, we emphasize that reflexive
motions are idling {\em at the level of abstraction of the automaton}.
When abstracting automata -- that is, constructing comparisons and
simulations -- we may have cause to abstract away internal
motions which are not germane to the analysis task at hand. Thus
reflexive motions may be thought of as representing motions unimportant
at the level of abstraction of the automaton, and not necessarily a strictly
idle state of the process being modelled.

\medskip

The proof of the following proposition is straightforward.
\begin{proposition}
\label{prop:simul_implies_appearance}
If there is a simulation $f\colon S \simto T$ then~$S$~and~$T$
have the same set of reduced appearances.
\end{proposition}

Of course, in the above proposition,  we are only considering
behaviours beginning at initial states.
For the connection between this
notion of simulation and  notions of observational
equivalences such as \cite{milner:comm_con}, 
the reader is referred to~\cite{JNW:bisim_open} and~\cite{CS:sync_async}.

The comparisons of
philosophers~$p\colon P' \compto P$, ~$q\colon P'' \compto P$
and~$r\colon P'' \compto P'$ mentioned above are examples of
simulations. One class of trivial (but, nevertheless important)
simulations is provided by the subautomata reachable from
the initial states. That is, for every~$S\colon X \to Y$,
the identity graph morphism of $\reachable{S}$ provides a
simulation~$1_{\reachable{S}}: S \simto S$.

\begin{proposition}
\label{prop:sim_deadlocks}
Suppose~$f\colon S \simto T$ is a simulation.
If~$v$ is a reachable deadlock of~$S$  
then~$f(v)$ is a reachable deadlock of~$T$. 
\end{proposition}

The above proposition indicates that simulations may be used for
detecting deadlocks:
given an automaton~$S$ we try to find 
a simulation~$f\colon S \simto T$
where~$\reachable{T}$ has significantly less states
than~$\reachable{S}$ (that is,~$\reachable{T}$ is a quotient
of~$\reachable{S}$); we then look for deadlocks~$v$ in $\reachable{T}$;
if $\reachable{T}$ has no deadlocks, we conclude that neither does~$S$;
and if there are deadlocks~$v$ in~$\reachable{T}$, we check to see if there
are any among the states~$w \in f^{-1}(v)$ of $\reachable{S}$.
We give an example of this process at the end of this section.

\medskip

Action sets, automata and simulations also form
a discrete Cartesian bicategory.  The operations
composition, binding, product and feedback of comparisons
induce the same operations on simulations. 

\subsubsection{Composition}
\label{section:simulation_composition}

Given simulations~$f\colon R \simto S$ and~$g\colon S \simto T$
where~$R$,$S$, and~$T \colon X \to Y$ are automata,
there exists a simulation~$\compose{g}{f}\colon R \simto T$
called the \termdef{composite} of~$f$~and~$g$. 

The composite is formed by composing the
comparisons~$\reachable{R} \compto \reachable{S}$
and~$\reachable{S} \compto \reachable{T}$ being the data provided for~$f$
and~$g$.  The requisite lifting property is easily established by first
lifting via~$g$, and then lifting each component of this lifting via~$f$.

As was the case with comparisons, for any two action sets~$X$~and~$Y$ we
can form a category; namely the category~${\bf Sim}(X,Y)$ whose objects
are automata with left boundary~$X$ and right boundary~$Y$ and whose
arrows are simulations between these automata.

\subsubsection{Binding, feedback and product}
\label{section:simulation_bfp}

Given automata~$S$ and~$T \colon X \to Y$ and
automata~$U$ and~$V \colon Y \to Z$, together
with simulations~$f \colon S \simto T$ and~$g \colon U \simto V$, we
define a simulation~$\bind{f}{g}\colon \bind{S}{U} \simto \bind{T}{V}$
called the \termdef{binding} of~$f$ and~$g$.

Given a state~$(v,w)$ of~$\reachable{\bind{S}{U}}$, it is clear that~$v$
is reachable in~$S$ and~$w$ is reachable in~$U$.  Applying~$f$ to~$v$
and~$g$ to~$w$ thus yields a pair of states, and the reachability of~$(v,w)$
implies this image pair is reachable in~$\bind{T}{V}$.  This defines the
state map of~$\bind{f}{g}$.  The motion map is similarly obtained from~$f$
and~$g$.

The lifting property is obtained by lifting componentwise. Without loss
of generality, the lifted paths have the same length (if not, extend the
shorter path by prepending reflexive motions). All but the last
motion in each lifting has a reflexive image, and thus will be a motion
of~$\bind{S}{U}$. The images of the final motion in each lifting perform
a common action on the boundary~$Y$, since we are lifting a motion
of~$\bind{T}{V}$. Thus, since~$f$ and~$g$ are comparisons, the
lifted motions also agree on their actions on the common boundary.

\medskip

Given automata~$S$ and~$T \colon X \times Y \to Y \times Z$ and a
simulation~$f\colon S \simto T$, we shall construct a
simulation~$\fb{Y}{f}\colon \fb{Y}{S} \simto \fb{Y}{T}$
called the \termdef{feedback} of~$f$.

Given a state~$v$ of~$\reachable{\fb{Y}{S}}$, we have a path from the initial
state of~$S$ to~$v$ consisting only of motions performing the same action
on the two boundaries of type~$Y$. Such~$v$ is clearly a state
of~$\reachable{S}$, and applying~$f$ then gives us a similar
path in~$T$, and we see that~$f$ induces a comparison as required.

As in the case for binding, the lifting property follows from the property
for~$f$ together with the fact that motions with reflexive image clearly
perform the same action on the two boundaries of type~$Y$, and the final
lifted motion agrees after application of~$f$ and hence before it, since~$f$
respects the actions on boundaries.

\medskip

Given automata~$S$ and~$T \colon X \to Y$
and automata~$U$ and~$V \colon W \to Z$, together
with simulations~$f \colon S \simto T$ and~$g \colon U \simto V$,
we define
the \termdef{product} of~$f$ and~$g$, a
simulation~$\product{f}{g}\colon \product{S}{U} \simto \product{T}{V}$.

Observe
that~$\reachable{\product{S}{U}} = \product{\reachable{S}}{\reachable{U}}$,
and thus the product of the comparisons underlying the simulations~$f$ and~$g$
yields a comparison to underly~$\product{f}{g}$. The lifting is performed
componentwise, extending the shorter path by prepending reflexive motions
if required.

\bigskip

It is a crucial aspect of the theory that the operations on designs lift
to operations on simulations.  Thus, given a design, we can abstract parts
of the design (i.e. simulate them with simpler systems) and produce
abstractions of the whole system.  In the next section we shall indicate how
this can be used to support model checking in the concrete example of the
dining philosophers.

\subsection{Simulations and Dining Philosophers}
\label{section:sim_checking}
  
We now indicate how simulations may facilitate the task
of model checking.

\bigskip

Consider the two alternative dining philosophers presented
at the end of section~\ref{section:misa_dp} (figures~\ref{fig:nondet_phil}
and~\ref{fig:slow_phil}).  As noted there, the sizes
of the state spaces of these systems which are explored by minimal
introspective subsystem analysis grow exponentially with the number
of philosophers~$n$.

Using these alternative philosophers with the design of the usual dining
philosopher systems, we may construct
corresponding
systems~$\fb{L}{(\bind{P'}{Q})^n}$ and~$\fb{L}{(\bind{P''}{Q})^n}$.
We have already noted,
however, that there are two simulations~$p\colon P' \simto P$
and~$q\colon P'' \simto P$. Thus using the operations of
section~\ref{section:simulation_bfp} we can construct
simulations 
\[
\tilde{p} = \fb{L}{(\bind{p}{1_{\reachable{Q}}})^n}\colon
	\fb{L}{(\bind{P'}{Q})^n} \simto \fb{L}{(\bind{P}{Q})^n}
\]
and
\[
\tilde{q} = \fb{L}{(\bind{q}{1_{\reachable{Q}}})^n}\colon
	\fb{L}{(\bind{P''}{Q})^n} \simto \fb{L}{(\bind{P}{Q})^n}.
\]
 
Now apply the minimal introspective subsystem analysis to
the standard philosopher system (recall the explored state space
grows only quadratically with the number~$n$). This analysis
will find the unique deadlock~$d$ of $\fb{L}{(\bind{P}{Q})^n}$.
Recall from section \ref{section:misa_dp} that this state~$d$ corresponds to
each fork being in state~$r$ and each philosopher being in state~$1$.

We now know that the only deadlocks of the alternative
philosopher systems are contained in $\tilde{p}^{-1}(d)$
and  $\tilde{q}^{-1}(d)$. It is easy to calculate  these
sets of states -- for example,  a state in $\tilde{p}^{-1}(d)$
corresponds to each fork being in state $r$ and
each philosopher being in state $1$ or $1'$. In fact,
each $v \in \tilde{p}^{-1}(d)$ and each $w \in \tilde{q}^{-1}(d)$
is a deadlock of $\fb{L}{(\bind{P'}{Q})^n}$
and $\fb{L}{(\bind{P''}{Q})^n}$ respectively,
and these are the only deadlocks of these systems. 

\bigskip

What if we want to analyse the dining philosopher system for
arbitrary~$n$?
With the use of software tools (such a tool is
currently being specified and prototyped by the authors),
it is reasonably straightforward
to construct an automaton~$R: L \to L$, together with
a pair of simulations~$f_2 \colon (\bind{P}{Q})^2 \simto R$
and~$f' \colon \bind{\bind{P}{Q}}{R} \simto R$.

The compositionality of simulations
allows us to deduce that for any~$n \geq 2$, there is a
simulation~$f_n\colon (\bind{P}{Q})^n \simto R$. We define, inductively,
for~$n \geq 2$
\[
f_{n+1} \colon (\bind{P}{Q})^{n+1} = \bind{\bind{P}{Q}}{(\bind{P}{Q})^n}
	\simto \bind{\bind{P}{Q}}{R}
	\simto R
\]
where the first simulation is~$\bind{1_{\reachable{\bind{P}{Q}}}}{f_n}$ and
the second simulation is~$f'$.

In other words (from the the point of
view of observational equivalence and checking for deadlocks) 
we can replace a composed sequence
of philosophers and forks of any length by the simple system $R$.

In the case of checking the dining philosopher system for deadlocks, we first
form the
simulation~$g = \fb{L}{f_n}\colon \fb{L}{(\bind{P}{Q})^n} \simto \fb{L}{R}$
and then note that the automaton~$\fb{L}{R}$ has a unique deadlock~$c$.
It is easy to check that the only vertex~$v$ of~$\fb{L}{(\bind{P}{Q})^n}$
with the property that~$g(v) = c$ is that corresponding to each philosopher
being in state~$1$ and each fork in state~$r$, allowing us to conclude that
this is the only deadlock of~$\fb{L}{(\bind{P}{Q})^n}$.

%% file: examples.tex
In this section we shall present two further examples of systems
composed from automata with boundary. We model a scheduler, which is
responsible for ensuring certain execution order properties in a
collection of concurrent systems.
We also present a model of processes communicating via a channel, and indicate
how communication protocols may be modelled as systems of automata with
boundary. The goal of this section is not to present any deep insights into
the systems we model, but to demonstrate the expressive power of the
methodology, and the process of design within the methodology.

\subsection{Scheduling}
\label{section:schedule}

We have in mind a system which controls the execution of a number
of processes in order to meet certain specifications. Each process
has a certain part of its execution, called the {\em controlled}
section, which is of interest to the scheduler. This system
is also used as an example of the calculus described
in~\cite{milner:calc_conc}.

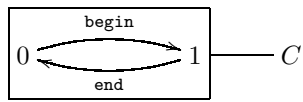
\begin{figure}[th]
\[
\xymatrix
{
0 \ar@/^/[rr]^{\kbegin}="top" &&
1 \ar@/^/[ll]^{\kend}="bot" & C
\save
	"1,1"."top"."bot"."1,3"*[F-]\frm{}
	\ar@{-}"1,4"
\restore
}
\]
\caption{\label{fig:process} A process~$P$ with a controlled section}
\end{figure}

Our processes~$P_1$,~$\ldots$,~$P_n$ each have one boundary, over
which they will
communicate with the scheduler.  We shall write~$C=\{\refl,\kbegin,\kend\}$
for the action set of these boundaries -- the process synchronizes
with a~$\kbegin$ to indicate it is entering its controlled section, and
synchronizes with a~$\kend$ to indicate it is leaving its controlled section.
Every behaviour of
each process must alternate~$\kbegin$ and~$\kend$ actions on its boundary.
We shall model the processes then with an automaton as shown in
figure~\ref{fig:process}.

In light of proposition~\ref{prop:sim_deadlocks} and
the constructions of section~\ref{section:simulation_bfp},
any analysis we perform
on systems involving~$P$ for deadlock will lift to the corresponding
systems using more complex processes, provided only that these processes are
simulated by~$P$.

In fact, one could argue that the property of being
simulated by~$P$ can be taken as a definition of the kind of process
we are interested in -- for the existence of such a simulation says
precisely that the process has states of two kinds (those mapping to 0
and those mapping to 1), and that it transits between these states by actions
observable on the boundary as $\kbegin$ and $\kend$.

\subsubsection{Design of the Scheduler}
\label{section:scheduler_design}

A scheduler~$S$ for~$n$ processes of this kind has~$n+1$ boundaries, each of
type~$C$. The scheduler will be connected to the processes~$P_i$
on~$n$ of its boundaries, with the last boundary being reserved for
control of the scheduler -- allowing an external agent to start and stop
the scheduler.

We shall construct a scheduler which forces the processes to~$\kbegin$
their controlled sections in a fixed, cyclic, order. That is to say,
we wish to control the processes such that in a global behaviour of the
system, the first nontrivial local motion of one of the process
components~$P_i$ is \kbegin\ by~$P_1$. The next nontrivial local motion
by a process component is \kbegin\ by~$P_2$, and so on.

Following the design of~\cite{milner:calc_conc}, we shall
construct the scheduler from a number of smaller
automata.  We shall use $n$~copies of an automaton~$N$, called a {\em notifier}.
Each notifier starts a single process and records the completion of its
controlled section. We also have a single master automaton~$M$, which responds
to outside control. The notifiers will pass a token around a circle.
When receiving the token, a notifier ensures its process~$\kbegin$s
its controlled section, and then passes the token on. The master automaton
hands the token out when it~$\kbegin$s, and will only~$\kend$ when it
holds the token, at which point it will not pass it on until
another~$\kbegin$ action occurs.

\begin{figure}[th]
\[
\xymatrix
{
&& 0 \ar[rr]^{(\kgo|-|-)}="top" & & 1 \ar[dd]^{(-|\kbegin|-)}="right" \\
G &&  & 2 \ar[ur]|{(-|\kend|-)} & && G\\
&& 3\ar[uu]^{(-|\kend|-)}="left" \ar[ur]|{(\kgo|-|-)} &
	& 4 \ar[ll]^{(-|-|\kgo)}="bot" \\
&&& C \\
\save
	"2,4"."top"."left"."bot"."right"*[F-]\frm{}
	\ar@{-}"2,1" \ar@{-}"2,7" \ar@{-}"4,4"
\restore
}
\]
\caption{\label{fig:notifier} A notifier~$N$}
\end{figure}
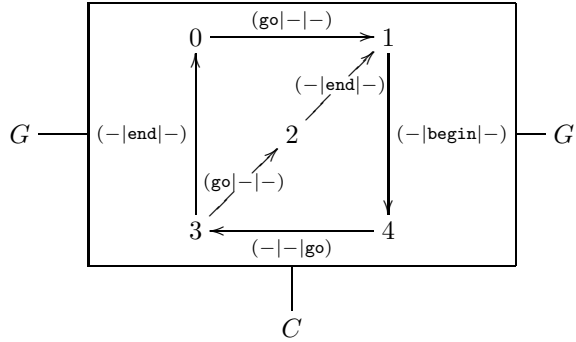

Each notifier is
a copy of an automaton~$N$, which has a boundary of type~$C$, and
two boundaries of type~$G=\{\refl,\kgo\}$ over which they will synchronize
with each other.  The notifier is shown in figure~\ref{fig:notifier} --
and edge labelled~$(a|b|c)$ indicates the action~$a$ on the left boundary~$G$,
the action~$b$ on the lower boundary~$C$, and the action~$c$ on the right
boundary~$G$.

Beginning in state~$0$, a notifier waits for a~$\kgo$ from its
left hand side.  It~$\kbegin$s its process, and then passes the~$\kgo$
to its right side. At this point, it waits to synchronize with
a~$\kgo$ from its left and a~$\kend$ from its process, before allowing
the process to start again.

\begin{figure}[th]
\[
\xymatrix
{
&& 0 \ar[rr]^{(-|\kbegin|-)}="top" & & 1 \ar[dd]^{(-|-|\kgo)}="right" \\
G &&  &  & && G\\
&& 3\ar[uu]^{(-|\kend|-)}="left" \ar@/^/[rr]^{(-|-|\kgo)}&
	& 2 \ar@/^/[ll]^{(\kgo|-|-)}="bot" \\
&&& C
\save
	"2,4"."top"."left"."bot"."right"*[F-]\frm{}
	\ar@{-}"2,1" \ar@{-}"2,7" \ar@{-}"4,4"
\restore
}
\]
\caption{\label{fig:master} The master~$M$}
\end{figure}
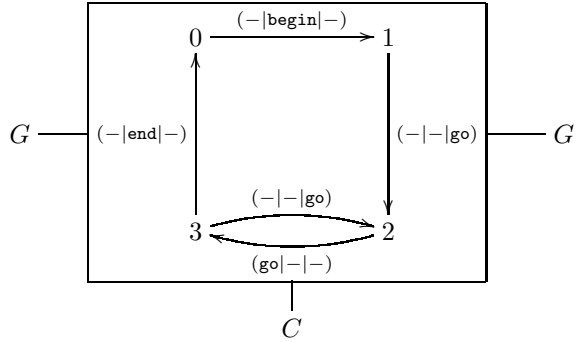

The master~$M$ likewise has a boundary of type~$C$, and
two boundaries of type~$G$.  It is shown in figure~\ref{fig:master},
with the same labelling convention as the notifier.

When the master receives a~$\kbegin$ on its control boundary~$C$, it
sends a~$\kgo$ to the right, and then waits for a~$\kgo$ on its left.
After receiving the~$\kgo$, it either passes it on or~$\kend$s, at which
point it must~$\kbegin$ again before passing~$\kgo$ to its right.

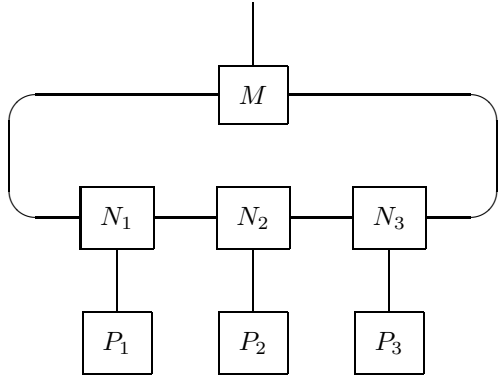
\begin{figure}[th]
\[
\xymatrix
{
&&&& \\
&& \boxedvar{M} \ar@{-}[u]
	\ar@{-}`r[rrd]`[rd][rd]
	\ar@{-}`l[lld]`[ld][ld]
	&& \\
& \boxedvar{N_1} \ar@{-}[d] \ar@{-}[r]
	& \boxedvar{N_2} \ar@{-}[d] \ar@{-}[r]
	& \boxedvar{N_3} \ar@{-}[d] & \\
& \boxedvar{P_1} & \boxedvar{P_2} & \boxedvar{P_3} & \\
&&&&
}
\]
\caption{\label{fig:scheduler_design}
	The design of a three process scheduler}
\end{figure}

The scheduler proper is then constructed by composing $n$~notifiers
and 1~master in a cycle.  The resultant automaton has~$n+1$ boundaries of
type~$C$. The $n$~boundaries arising from the notifiers are connected
to the processes we wish to control. The remaining boundary, arising
from the master, is connected to the external control mechanism.
Figure~\ref{fig:scheduler_design} shows the design of the final composite
in the case~$n = 3$.

In order to analyse this system below, we shall close the final boundary
by binding a controlling process (an automaton of the form of~$P$) to
the remaining boundary of this design.

\subsubsection{Analysis of the Scheduler}
\label{section:scheduler_analyze}

We shall now briefly indicate how one could analyze the scheduler system
using the methodology described in this paper.

Consider the design of the system as shown in figure~\ref{fig:scheduler_design},
and with an additional controlling process~$P$ bound to the remaining
boundary.

We begin by evaluating the binding~$\bind{N}{P}$.  We shall not draw this
binding in full, but we note that it is simulated by the automaton~$Q$ shown
in figure~\ref{fig:abstract_NP}.

\begin{figure}[th]
\[
\xymatrix
{
G & 0 \ar@/^/[rr]^{(\kgo|-)}="top" &&
1 \ar@/^/[ll]^{(-|\kgo)}="bot" & G
\save
	"1,2"."top"."bot"."1,4"*[F-]\frm{}
	\ar@{-}"1,5"
	\ar@{-}"1,1"
\restore
}
\]
\caption{\label{fig:abstract_NP} An automaton~$Q$ simulating~$\bind{N}{P}$}
\end{figure}

Thus the binding~$\bind{N}{P}$ is simulated by a system which
performs the~$\kgo$ action on its boundaries alternately. The image
on the initial state under the simulation is the state~$0$ in
the automaton~$Q$.

Now consider the master system.  When bound with a controlling process of the
form of~$P$, the resultant system is also simulated by~$Q$, although
this time with initial state having image~$1$ in~$Q$.

Thus the evaluation of the system of figure~\ref{fig:scheduler_design} is
simulated by a ring of~$n+1$ copies of~$Q$, with initial state having precisely
one copy of~$Q$ in the local state~$1$.  It is easy to see the evaluation of
such a system is isomorphic to the automaton with states length~$n+1$ cyclic
strings of~$0$s and~$1$s, and motions being those which replace a~$10$ substring
with a~$01$ substring.  Consider the subautomaton reachable from an
initial state being a cyclic string containing precisely 1 local state of~$1$:
This is clearly a (graph theoretic) cycle of~$n+1$ states.
Hence the system is deadlock free.

By keeping track of the simulations indicated in the preceding discussion,
one can deduce more from the constructed simulation, including the desired
behaviour in terms of the order of entering the controlled section for each
controlled process.

\subsection{Communication Protocols}
\label{section:comms}

\subsubsection{Notation}

For this section, we shall introduce an abbreviated notation
for drawing automata which allows for easier depiction of automata where
many states are similar.

As described here, the notation merely gives a compact representation of
certain automata of interest.  However, the authors intend to more fully
explore this notation, with a view to allowing specification of automata
using abstract data types via the interpretations
of \cite{walters:data_types} and \cite{gates:phd}.

Let us consider an automaton~$X \to Y$.
Our pictures will be graphs, with the following additional data:
\begin{enumerate}
\item
	associated to each vertex is a given a set~$V$. We typically abuse
	notation by denoting and referring to the vertex as~$V$,
	provided no confusion arises.
\item
	associated to each edge~$V \to W$ is a subset
	of~$X \times V \times W \times Y$. We shall denote the subset by
	a label~$(x(i)|v(i)\to w(i)|y(i))$ where~$i$ ranges over some
	(typically implicit) indexing set.
\end{enumerate}
An automaton is associated with such a graph as follows:
A vertex denoted~$V$ indicates a set of states indexed by~$V$;
an edge labelled~$(x(i)|v(i)\to w(i)|y(i))$ indicates a
family of motions~$v(i) \to w(i)$ labelled~$x(i)$ on the
boundary~$X$ and~$y(i)$ on the boundary~$Y$.

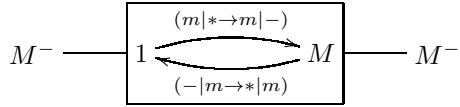
\begin{figure}[th]
\[
\xymatrix
{
M^{\refl}
& 1 \ar@/^/[rr]^{(m|\ast\to m|\refl)}="top" &&
	M \ar@/^/[ll]^{(\refl|m\to\ast|m)}="bot"
& M^{\refl}
\save "1,2"."top"."bot"."1,4"*[F-]\frm{} \ar@{-}"1,1" \ar@{-}"1,5" \restore
}
\]
\caption{\label{fig:message_passer} A message passer automaton}
\end{figure}

For example, given a set~$M$ of messages, let us write~$M^{\refl}$
for the boundary obtained by adjoining a trivial action to~$M$.
Figure~\ref{fig:message_passer} shows an
automaton with boundaries~$X = M^{\refl}$ and~$Y = M^{\refl}$. The
automaton has~$M + 1$ states, and nontrivial motions of two kinds:
\begin{enumerate}
\item
	from the lone state of~$1$ to each state of~$M$, this motion
	being labelled by the target state on the left boundary, and~$\refl$
	on the right boundary,
\item 
	from each state of~$M$ to the lone state of~$1$, this motion
	being labelled by~$\refl$ on the left boundary and the source state
	on the right boundary.
\end{enumerate}
Such an automaton is a simplistic delayed message passer -- it synchronizes
with its left boundary to obtain~$m \in M$ (storing it internally by
moving to an appropriate state), and then synchronizes with its right
boundary to pass~$m$ on (and forgetting the~$m$ in the process).
Considering the definition of binding of automata, we note that
this automaton is incapable of losing messages -- bound systems
synchronize on the boundary actions which represent passing/receiving
a message to/from the automaton.

\subsubsection{Channels}

By a {\em channel~$C$ of type~$M$}, for a given set of messages~$M$,
we mean an automaton~$C \colon M^{\refl} \to M^{\refl}$.
We have in  mind that an action~$m$ being performed on the left boundary is
sending the message~$m$ down the channel, and some time later the right
boundary will perform the action~$m$ as the message emerges.  However,
we do not require these properties of a channel, as we wish to model
channels which may lose, modify or reorder messages.

Given the above
discussion of the message passer, we note that a synchronization with
a channel is considered to be a tightly coupled interaction whereby
the channel accepts a message from its boundary.  Synchronization occurs
when a message is transferred across the boundary. That is, our I/O is
fundamentally blocking I/O.

Non-blocking I/O is modelled by having a
automaton which can receive messages in any (or almost any) state.
We note that this is an accurate model of non-blocking I/O.
Such I/O is not distinguished in that it does not synchronize, but
rather in that it synchronizes {\em locally} -- that is, with lower layer
processes in the local communication library rather than with a distant
system.

To reconcile the tightly coupled nature of the synchronization in
the binding operation with our desire to model channels which lose messages,
we construct channels which literally lose messages -- it is a property
of the channel that a message which enters it may not emerge. 
By explicitly modelling that part of the system that loses messages,
we can provide precise analyses of whether or not certain protocols
lose messages.

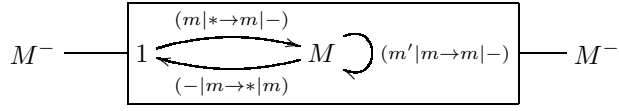
\begin{figure}[th]
\[
\xymatrix
{
M^{\refl} & 1 \ar@/^/[rr]^{(m|\ast\to m|\refl)}="top" &&
M \ar@/^/[ll]^{(\refl|m\to\ast|m)}="bot"
  \ar@(ur,dr)[]^{(m'|m\to m|\refl)}="right"
&&& M^{\refl}
\save
	"1,2"."top"."bot"."right"*[F-]\frm{} \ar@{-}"1,1" \ar@{-}"1,7"
\restore
}
\]
\caption{\label{fig:cap1_channel} A channel of capacity 1}
\end{figure}

Such
a channel is shown in figure~\ref{fig:cap1_channel}. This is a channel
of type~$M$, which we shall refer to as a capacity~$1$ channel.  If
the channel is empty (in the state~$1$), an input transition of~$m$
results in the message~$m$ being stored (in one of the states~$M$).
This can later be read by an output transition, and the channel
returns to the empty state.  Any input messages supplied to the
channel while it is full are simply lost.

Precisely speaking, we may consider any sequence~$\sigma$ of actions on the
left boundary of the automaton.  For any behaviour~$\beta$ of the
automaton with reduced appearance on the left boundary being the given
sequence~$\sigma$, the reduced appearance on the right boundary is
a subsequence of~$\sigma$.  Further, there exists a behaviour~$\beta$
for which the reduced appearance on the right boundary is precisely the
sequence~$\sigma$.

\subsubsection{Protocols}

By a {\em protocol~$P$ of type~$M$ implemented on a channel of type~$N$},
we mean a pair~$(S,R)$ of automata, called the
{\em sender}~$S \colon M^{\refl} \to N^{\refl}$ and the
{\em receiver}~$R \colon N^{\refl} \to M^{\refl}$.
Given a channel~$C$ of type~$N$, we can construct the
channel~$\bind{\bind{S}{C}}{R}$.

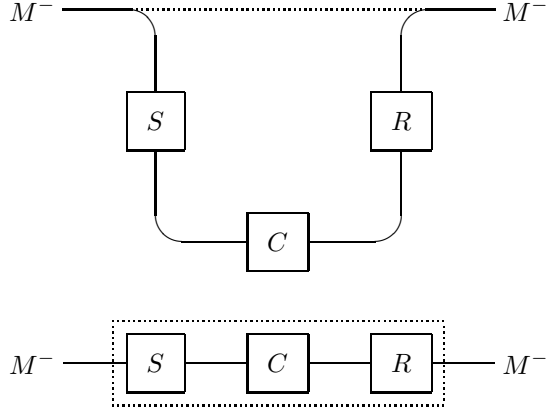
\begin{figure}[th]
\[
\xymatrix
{
M^\refl
	\ar@{-}`r[dr][dr]
	\ar@{.}[rrrr]
	&&&&
M^\refl
	\ar@{-}`l[dl][dl] \\
& \boxedvar{S} \ar@{-}`d[dr][dr]
&& \boxedvar{R} \ar@{-}`d[dl][dl] & \\
&& \boxedvar{C} &&  \\
M^\refl \ar@{-}[r]
	& \boxedvar{S} \ar@{-}[r]
	& \boxedvar{C} \ar@{-}[r]
	& \boxedvar{R} \ar@{-}[r]
	& M^\refl
\save "4,2"."4,4"*!<0.4em,0em>+<1em,1em>\frm{.} \restore
}
\]
\caption{\label{fig:virtual_channel} Virtual vs. designed channel}
\end{figure}

This latter channel is
what is usually termed the {\em virtual channel} provided by the
protocol.  Given the definitions of this paper, it is in fact a
channel, no less real for the fact it is built from simpler automata.
The authors suggest the term ``designed channel'' to distinguish the
latter channel from the former.
A composite of this kind is shown in figure~\ref{fig:virtual_channel} --
the term virtual channel arises by thinking of the dotted line in the
upper diagram as a direct connection; the author's point of view
is that the dotted box in the lower diagram shows a designed channel
constructed from the protocol and underlying channel.

One goal of protocol design is to construct
the automata~$S$ and~$R$ in such a way that this virtual channel has
better properties than the underlying channel~$C$.  Typically, we wish
to show that given certain properties of the channel~$C$, the virtual
channel~$\bind{\bind{S}{C}}{R}$ has certain other properties.

More generally, we may state the problem of protocols as follows:
Given a family of channels~$C_i$, a desired channel type~$M$, and
requirements on the behaviours of the desired virtual channel,
we need to construct automata~$S \colon M^{\refl} \to N^{\refl}$
and~$R \colon N^{\refl} \to M^{\refl}$ such that the
channel~$\bind{\bind{S}{C}}{R}$ has the desired
properties for some~$C \colon N \to N$ selected from our family~$C_i$.
Note that the family~$C_i$ models ``the sorts of channels available to
us at this level of abstraction'', and would typically be described as
the closure under certain operations of certain basic channels -- for
example, any channel which is a product of capacity 1 channels of any
type.

\subsubsection{Message Acknowledgement}
\label{section:ack}

Given that the capacity~1 channel can lose messages, we might ask to
establish a virtual channel solving this problem.  One solution is to
acknowledge sent messages.  We shall use a channel from the receiver to
the sender of type~$A = \{\ack\}$ to carry the acknowledgements.  That
is, we shall build a virtual channel of type~$M$ from a channel of
type~$N = M\times A + M + A$.  Note that the type~$N$ of this channel
should be thought of as the the product of the types~$M$ and~$A$ in
the sense that~$N^\refl = M^\refl \times A^\refl$.

The sender and receiver automata~$S$ and~$R$ are shown in
figures~\ref{fig:ack_sender} and~\ref{fig:ack_receiver} respectively.
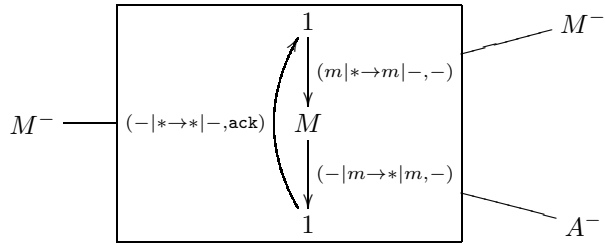
\begin{figure}[th]
\[
\xymatrix
{
&&& 1 \ar[d]^{(m|\ast\to m|\refl,\refl)} &&& M^\refl \\
M^\refl &&& M \ar[d]^{(\refl|m\to\ast|m,\refl)}="right" \\
&&& 1 \ar@/^3ex/[uu]^{(\refl|\ast\to\ast|\refl,\ack)}="left" &&& A^\refl
\save
	"left"."1,4"."3,4"."right"*[F-]\frm{}
	\ar@{-}"2,1" \ar@{-}"1,7" \ar@{-}"3,7"
\restore
}
\]
\caption{\label{fig:ack_sender}
	Sender~$S$ for Message Acknowledgement Protocol}
\end{figure}
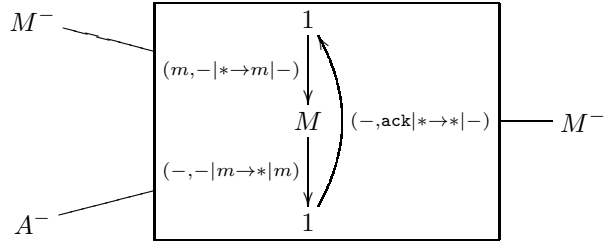
\begin{figure}[th]
\[
\xymatrix
{
M^\refl &&& 1 \ar[d]_{(m,\refl|\ast\to m|\refl)} \\
&&& M \ar[d]_{(\refl,\refl|m\to\ast|m)}="left" &&& M^\refl \\
A^\refl &&& 1 \ar@/_3ex/[uu]_{(\refl,\ack|\ast\to\ast|\refl)}="right" \\
\save
	"right"."1,4"."3,4"."left"*[F-]\frm{}
	\ar@{-}"1,1" \ar@{-}"3,1" \ar@{-}"2,7"
\restore
}
\]
\caption{\label{fig:ack_receiver}
	Receiver~$R$ for Message Acknowledgement Protocol}
\end{figure}

One can easily evaluate the binding~$\bind{S}{R}$
to determine the behaviour of the message acknowledgement protocol
over a perfect channel.

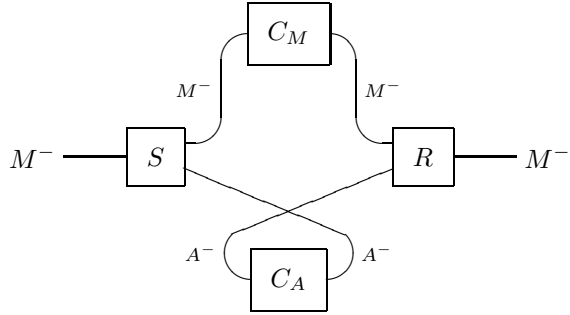
\begin{figure}[th]
\[
\xymatrix
{
&& \boxedvar{C_M}
	\ar@{-}`r[d]`[dr]!<0em,0.5em>^{M^\refl}[dr]!<0em,0.5em>
	\ar@{-}`l[d]`[dl]!<0em,0.5em>_{M^\refl}[dl]!<0em,0.5em>
	&&
	\\
M^\refl &
	\boxedvar{S}
	\ar@{-}[l]
	&&
	\boxedvar{R}
	\ar@{-}[r]
	& M^\refl 
	\\
&&
	\boxedvar{C_A}
	\ar@{-}`r[u]`^ul[ul]_{A^\refl}[ul]
	\ar@{-}`l[u]`_ur[ur]^{A^\refl}[ur]
	&&
}
\]
\caption{\label{fig:comms_design}
	The design of the message acknowledgement protocol}
\end{figure}

However, we wish to analyze the protocol over a pair of capacity 1
channels running in opposite directions -- a channel of type~$M$
from~$S$ to~$R$ and a channel of type~$A$ from~$R$ to~$S$.
The design of the system we wish to analyze is shown in
figure~\ref{fig:comms_design}.

Given an automaton~$S \colon X \to Y$, the
automaton~$S^{\rm op} \colon Y \to X$ is constructed by interchanging the
boundaries -- we may call~$S^{\rm op}$ as the \termdef{opposite}
of~$S$.  Thus the channel of interest in this context is the
product~$C_M \times C_A^{\rm op}$ of a pair of capacity~1 channels of
type~$M$ and and~$A$ respectively (running in opposite directions).
Note that design diagram of figure~\ref{fig:comms_design} need not
mention the opposite explicitly - the required connections being
expressed by appropriate wires.

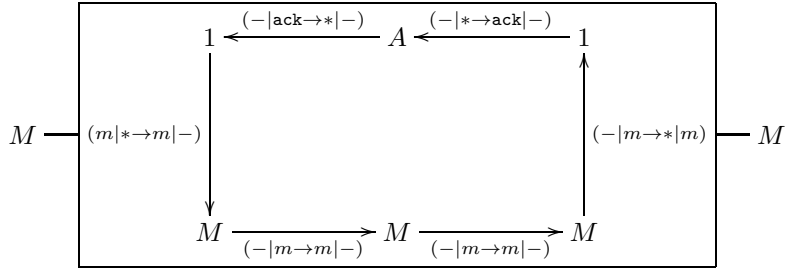
\begin{figure}[th]
\[
\xymatrix
{
&& 1 \ar[dd]_{(m|\ast\to m|\refl)}="left"
	&& A \ar[ll]_{(\refl|\ack\to\ast|\refl)}="top"
	&& 1 \ar[ll]_{(\refl|\ast\to\ack|\refl)}\\
M &&  &&&& && M \\
&& M \ar[rr]_{(\refl|m\to m|\refl)}="bot"
	&& M \ar[rr]_{(\refl|m\to m|\refl)}
	&& M \ar[uu]_{(\refl|m\to \ast|m)}="right"  \\
\save
	"left"."top"."right"."bot"*[F-]\frm{}
	\ar@{-}"2,1" \ar@{-}"2,9"
\restore
}
\]
\caption{\label{fig:MAP_cap1}
	Message Acknowledgement Protocol over capacity 1 channels}
\end{figure}

Our goal now is to explain why the channel so constructed meets the
design goals - that is, has only behaviours which have identical
reduced appearances on each boundary. We do this by evaluating the
design of figure~\ref{fig:comms_design} - the reachable part is shown in
figure~\ref{fig:MAP_cap1}.  It is clear that this channel is simulated
by the message passer of figure~\ref{fig:message_passer} - map the states
in the top row to the unique state of~$1$ in figure~\ref{fig:message_passer},
and map the states in the bottom row to the corresponding states of~$M$
in figure~\ref{fig:message_passer}.
Proposition~\ref{prop:simul_implies_appearance} of
section~\ref{section:simulations}
now provides the desired result.

\begin{figure}[th]
\[
\xymatrix
{
&& 1 \ar[dd]_{(m|\ast\to m|\refl)}="left"
	&& A \ar[ll]_{(\refl|\ack\to\ast|\refl)}="top"
	&& 1 \ar[ll]_{(\refl|\ast\to\ack|\refl)}
		\ar[lld]|{(\refl|\ast\to \ast|\refl)} \\
M &&  && 1 && && M \\
&& M \ar[rr]_{(\refl|m\to m|\refl)}="bot"
		\ar[rru]|{(\refl|m\to \ast|\refl)}
	&& M \ar[rr]_{(\refl|m\to m|\refl)}
	&& M \ar[uu]_{(\refl|m\to \ast|m)}="right"  \\
\save
	"left"."top"."right"."bot"*[F-]\frm{}
	\ar@{-}"2,1" \ar@{-}"2,9"
\restore
}
\]
\caption{\label{fig:MAP_lossy}
	Message Acknowledgement Protocol over lossy channels}
\end{figure}
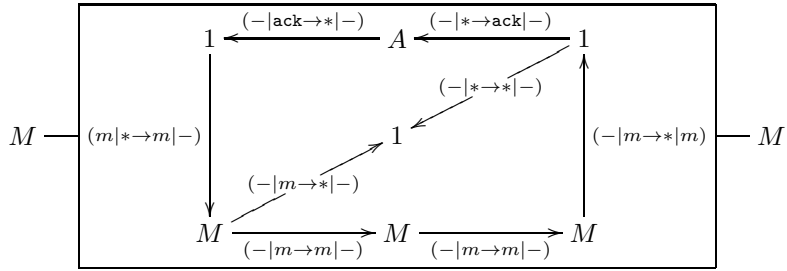

What happens if the channel of type~$M$ being used is not capacity~1, but
may in fact lose messages arbitrarily.  Such a channel would be modelled
by an automaton similar to that of figure~\ref{fig:cap1_channel}, but
with an additional transition labelled~$(m|\ast\to\ast|-)$ from the
state~$1$ to itself. One can readily evaluate the design of
figure~\ref{fig:comms_design} using this channel in place of~$C_M$,
and with a correspondingly modified channel in place of~$C_A$.
The result is the automaton of figure~\ref{fig:MAP_lossy}. In this
case the system deadlocks if a message or an acknowledgement is lost.

To repair this defect one typically uses timeouts and retransmission, but
the analysis of such protocols, while of direct interest to the authors,
is beyond the goals of the current paper.  It is observed however, that
timeouts could be modelled in the theory of automata with boundary by
using timeout motions in the sender and receiver automata.

%% file: conclusion.tex
We have presented the basic theory of automata with boundary,
together with examples designed to elucidate the presentation and
show the scope of the theory described here.  It is important to
reiterate that the theory provides an algebra for constructing systems
from primitive elements.
One of the crucial aspects of this theory is the attempt to capture the design of a system
as a precise theoretical element, distinct from the system itself
and its implementation.

The
underlying mathematical formalism of the approach has been explored in
~\cite{CW:cart_bicat},~\cite{katis:phd},~\cite{KSW:alg_feedback},%
~\cite{KSW:bicat_processes}, and~\cite{KSW:span_graph}, and the interested
reader is referred there. There is still work to be done in clarifying
some details of the mathematics appropriate to the model, for
example~\cite{GK:designs} and other papers in preparation by the authors.
We note also a precise description of the application of the bicategory
Span(Graph) to the domain of asynchronous circuit design is given in
~\cite{KSWW:cats_circ} and ~\cite{weld:phd}.

We have proposed an algorithm for model checking systems for deadlock
which fits comfortably with the theory, and illustrates the principle
that incorporation of designs as an element of the theory has
benefits in other areas.  As noted at the conclusion of
section~\ref{section:misa_dp}, this algorithm is simplistic and does not
always perform well -- more work in understanding the applications,
limitations and possible evolution of algorithms based on these ideas
is clearly warranted.

The theory supports abstraction of automata via the notions of comparison
and simulation. The algebra used to construct systems extends to an algebra
including the abstraction mechanisms, facilitating the construction of
abstractions of larger systems.  While the authors are still investigating
the use of this technique, some indication of the benefits this approach
yields are seen in section~\ref{section:sim_checking}: abstractions
may be used in conjunction with model checking to check larger systems; and
the compositionality of the abstractions allows theoretical checking of
families of systems.

In addition, the authors note that the combinatorial nature of the theory
presented here makes it ideal for machine manipulation.  As mentioned in
section~\ref{section:sim_checking} the authors are presently prototyping
tools designed to facilitate calculation in the algebra presented in this
paper.  It is hoped that such tools will allow calculation with larger models,
such as several layers of a multilayer network protocol, both to demonstrate
the applicability of the theory and to further refine the ideas presented
in the current work.